\documentclass{aa}[referee] 
\usepackage[varg]{txfonts}
\usepackage{amsmath}
\usepackage{amsfonts}
\usepackage{amssymb}
\usepackage{upgreek}
\usepackage{natbib}
\usepackage{xcolor}
\usepackage{hyperref}
\usepackage{graphicx}
\hypersetup{
    colorlinks=true,
    linkcolor=blue,
    citecolor=blue,
    urlcolor=blue,
    filecolor=magenta,      
}

\bibpunct{(}{)}{;}{a}{}{,} 

\begin{document}

\title{Numerical simulations of turbulence in prominence threads induced by torsional  oscillations}
\author{Sergio D\'{i}az-Su\'{a}rez \inst{\ref{inst1},\ref{inst2}} \and Roberto Soler \inst{\ref{inst1},\ref{inst2}}}
\institute{Departament de F\'{i}sica, Universitat de les Illes Balears, E-07122, Palma de Mallorca, Spain \label{inst1} \and Institute of Applied Computing \& Community Code (IAC3), Universitat de les Illes Balears, E-07122, Palma de Mallorca, Spain \label{inst2}; \email{s.diaz@uib.es}}

\date{Received 10 October 2023 /Accepted 23 December 2023}
\abstract
{Threads are the main constituents of prominences. They are  dynamic structures that display  oscillations, usually interpreted as magnetohydrodynamic (MHD) waves. Moreover, instabilities such as the Kelvin-Helmholtz instability (KHI) have also been reported in prominences. Both waves and instabilities may affect the thermodynamic state of the threads.}
{We investigate the triggering of turbulence in prominence threads caused by the nonlinear evolution of standing torsional Alfv\'{e}n waves. We study  the heating in the partially ionized prominence plasma as well as possible observational signatures of this dynamics.}
{We modeled a prominence thread as a radially and longitudinally nonuniform cylindrical flux tube with a constant  axial magnetic field embedded in a much lighter and hotter coronal environment. We perturbed the flux tube with the longitudinally fundamental mode of standing torsional Alfv\'{e}n waves. We numerically solved the three-dimensional (3D) MHD equations to study the temporal evolution in both ideal and dissipative scenarios. In addition, we performed forward modeling to calculate the synthetic H$\alpha$ imaging.}
{The standing torsional Alfv\'{e}n waves undergo phase-mixing owing to the radially nonuniform density. The phase-mixing generates  azimuthal shear flows, which eventually trigger the KHI and, subsequently, turbulence. When nonideal effects are included, the obtained plasma heating is very localized in an annulus region at the thread boundary and does not increase the temperature in the cool core. Instead, the average temperature in the thread decreases owing to the mixing of internal and external plasmas. In the synthetic observations, first we observe periodic pulsations in the H$\alpha$ intensity caused by the integration of the phase-mixing flows along the line of sight. Later, fine strands that may be associated with the KHI vortices are seen in the synthetic H$\alpha$ images.}
{Turbulence can be generated by standing torsional Alfv\'{e}n waves in prominence threads after the triggering of the KHI, although this mechanism is not enough to heat such structures. Both the phase-mixing stage and the turbulent stage of the simulated dynamics could be discernible in high-resolution H$\alpha$ observations.}


\titlerunning{Torsional oscillations in prominence threads}
\authorrunning{S. D\'{i}az-Su\'{a}rez \& R. Soler}
\maketitle

\section{Introduction}
Solar quiescent prominences are majestic structures of plasma sustained against gravity thanks to magnetic fields, whose strengths vary between 3 G and 60 G \citep[see, e.g.,][]{Mackay10,Gibson18}. Such structures can be seen as bright structures in the solar limb or as dark structures on the solar disk, in the case of which they are referred to as filaments that are formed in filament channels. Projected on the photosphere, solar prominences are near to a neutral line, which separates regions of opposite magnetic polarity \citep[][]{Babcock55,Howard64}. Typically, solar prominences have temperatures between 7500 K and 9000 K, gas pressure between 0.1 $ \mathrm{dyn/cm^{2}}$ and 1 $ \mathrm{dyn/cm^{2}}$, and number densities between  $ 10^{10}\;\mathrm{cm^{-3}}$ and $ 10^{11} \;\mathrm{cm^{-3}}$ \citep{Labrosse10,Parenti14,engvoldchapt2}. Moreover, solar prominences are usually suspended at heights that corresponds to the lower corona. In such environment, temperatures are above $10^{6} $ K, but the number density is low, $\sim 10^{8}\;\mathrm{cm^{-3}}$ \citep[see, e.g.,][]{Priest14}. Consequently, the plasma inside a prominence is cold and dense compared with the coronal plasma. Importantly, the plasma in prominences is only partially ionized \citep{Labrosse10}.

High-cadence and high-resolution  H$\alpha$ observations showed that  prominences are formed by a myriad of substructures called prominence threads \citep{Lin05,Lin07,Lin08}. Prominence threads are quite thin $ (\sim 0.21 $ Mm) and long (between $ \sim 3.5 $ and 28 Mm) structures, which are believed to be embedded in much longer magnetic tubes \citep[see, e.g.,][]{Lin05,Lin07} of lengths of the order of 100 Mm or more whose feet are rooted in the lower atmosphere \citep{Terradas08thr}. Consequently, prominence threads are usually modeled as thin magnetic flux tubes with their ends fixed at the photosphere \citep{Ballester89,Rempel99,Soler12,Andreuu21,Melis23}.

Prominence threads are so dynamic that their continuous presence in H$\alpha$ observations is typically short. It ranges between few minutes and 20 minutes \citep{Lin05}. Magnetohydrodynamic (MHD) waves and oscillations play a predominant role in the dynamics of threads \citep[see][]{Arregui18LRSP}. There are many reports of transverse oscillations and propagating waves in prominence threads with periods between 2 min and $ \sim $ 15 min \citep{Lin07,Lin09,Ning09,Orozco14,Li22}. These oscillations are interpreted as kink MHD waves \citep{Terradas08thr,Lin09,Soler11,Okamoto15,Li22}. The driver of these waves is believed to be the convective motions at the photosphere, where the prominence magnetic field is anchored \citep{hillier2013}.

The present paper deals with torsional Alfv\'{e}n waves, which are transverse MHD waves that periodically twist the magnetic field lines in a flux tube. It has been postulated that they might play a role in the heating of the solar corona \citep{Hollweg78,Cranmer05,Cargill11,Mathioudakis13,Soler19apj,Tom2020,Nakariakov2020} and in the acceleration of the solar wind \citep{Charbonneau95,Cranmer09,Matsumoto12,Shoda18}. In straight flux tubes with a purely axial magnetic field, these waves can maintain their pure magnetic nature despite inhomogeneities \citep{Goossens11}. Torsional Alfv\'{e}n waves are nearly incompressible and are polarized perpendicularly to the magnetic field lines producing periodic perturbations in the azimuthal components of the magnetic field and the velocity.  Recently, \citet{Soler19apj,Soler21APJ} found that torsional Alfv\'{e}n waves can transport large energy fluxes when they propagate from the photosphere to a coronal loop despite the filtering role of the chromosphere. They found that the energy flux is channeled at the frequencies that match the natural frequencies of the coronal loop, generating global standing torsional oscillations. The generation of standing modes is possible because coronal loops can act as Alfv\'{e}n cavity resonators \citep{Hollweg84,Hollweg84b}. However, the energy distribution is not equally distributed among the different eigenmodes, with the fundamental mode being the major contributor. 

To the best of our knowledge, there is no direct evidence of torsional Alfv\'{e}n waves in prominence threads, although there have been direct reports of rotational motions caused by magnetic reconnection \citep{Okamoto16}. Nonetheless, \citet{Jess09} detected torsional Alfv\'{e}n waves at photospheric bright points through spectral line nonthermal widening \citep[see][]{Zaqarashvili2003}. \citet{Morton13} reported the presence of torsional Alfv\'{e}n and kink waves in chromospheric swirls. \citet{DePontieu12,DePontieu14} detected torsional motions in spicules that could be compatible with torsional Alfv\'{e}n waves. This type of waves  were also reported in the chromosphere by \citet{Srivastava17b}. More recently, \citet{Kohutova20} detected torsional Alfv\'{e}n waves at coronal heights using the  Interface Region Imaging Spectrograph (IRIS; \citeauthor{Depontieu14iris}~\citeyear{Depontieu14iris}). Moreover, some oscillations generated during solar flares can be interpreted as torsional Alfv\'{e}n waves \citep{Aschwanden20} and torsional Alfv\'{e}n waves have been reported in the solar wind \citep{Raghav18}. Therefore, in view of their ubiquity, it is likely that torsional Alfv\'{e}n waves are also present in prominences, although their direct detection remains elusive.

In a line-tied flux tube that is transversely nonuniform, either in density or in magnetic field, there is an Alfv\'{e}n frequency continuum \citep[see][]{Halberstadt93}. Consequently, Alfv\'en modes at different radial positions in the flux tube become out of phase as time increases. This phenomenon is the well-known process of phase-mixing \citep[see, e.g.,][]{HeyvaertPriest83,Nocera84,Moortel00,Smith07,Prokopyszyn2019,Diazsoler21b}. Phase-mixing is a linear process that generates shear flows perpendicularly to the magnetic field lines and transports Alfv\'en wave energy from large to small  perpendicular scales. The rhythm at which the energy is transported depends on the gradient of the Alfv\'en frequency \citep[see][]{Mann95}. Eventually, the phase-mixing shear flows trigger the Kelvin-Helmholtz instability (KHI) as \citet{HeyvaertPriest83} and \citet{Browning84} analytically predicted. This result is numerically confirmed in \citet{Guo19} and \citet{diazsoler21aanda}. Although with some important differences, an equivalent dynamics also appears in simulations of kink waves \citep[see, e.g.,][]{Terradas08,Terradas18,Antolin14,Magyar16,Howson17bres,Howson17a,Karampelas17,Karampelas18,Karampelas19,Guo19,Pascoe20,Shi21,Magyar22}. The KHI generates vortices that nonlinearly break into smaller and smaller vortices leading naturally to turbulence. Although not related to the Alfv\'{e}n waves, there are observations of the  KHI in coronal mass ejections \citep{Foullon11} and in prominences \citep{Berger17,HillierPolito18,Yang18}. In the latter case, turbulence is also reported  \citep[see][]{Leonardis12}.

In the present work we study the nonlinear evolution of  torsional Alfv\'{e}n waves in quiescent prominence threads. The purpose of this investigation is twofold. On the one hand, we aim to the explore the process of turbulence generation mediated by torsional waves. Such a mechanism has been studied before in the case of coronal loops \citep{diazsoler21aanda,Diazsoler22} but, to the best of our knowledge, not in prominence threads.  Unlike the fully ionized coronal loops,  prominence threads are only partially ionized. In the partially ionized prominence plasma, ambipolar diffusion and, to a lesser extent, Ohmic diffusion, are important dissipation mechanisms \citep{Khomenko14b,Ballester18,Melis21}. Consequently, ambipolar and Ohmic diffusion are included here as nonideal effects that can dissipate Alfv\'{e}n waves and the associated turbulence, and so potentially heat prominence threads. On the other hand, we aim to explore what kind of signatures the dynamics of torsional oscillations may leave in synthetic H$\alpha$ observations, so that  observers may look for those signatures in real observations in order to detect the still elusive torsional waves in prominences.

This paper is organized as follows. In Sect. \ref{sec:setup}, we describe the numerical setup. The results from both ideal MHD and nonideal MHD simulations are given in Sect. \ref{sec:simulations}. The synthetic H$\alpha$ observations are presented and analyzed in Sect. \ref{sec:forward}. Finally, we discuss the conclusions of this work in Sect. \ref{sec:conclusions}.


\section{Numerical setup}
\label{sec:setup}

\subsection{Prominence thread model}
We represent a prominence thread as a straight magnetic flux tube of length, $L,$ and radius, $R,$ permeated by  a uniform longitudinal magnetic field, namely, $ \mathbf{B}=B_{0} \hat{z}$, where $B_0 = 5.2$~G throughout. We consider $R=$~1~Mm and $ L = 50 R $, so that $L=$~50~Mm.  The value of $L$ used here is shorter than that usually reported from observations by a factor of 2, approximately. \citet{Terradas08thr} inferred the minimum length of the magnetic tube of an active-region prominence thread reported by \citet{Okamoto07} as $L\approx$~104.8 Mm. The reason for considering a shorter tube is to speed up the computations, since the periods of the standing waves are proportional to $L$. 

\begin{figure*}[htbp!]
\resizebox{0.75\hsize}{!}{\includegraphics{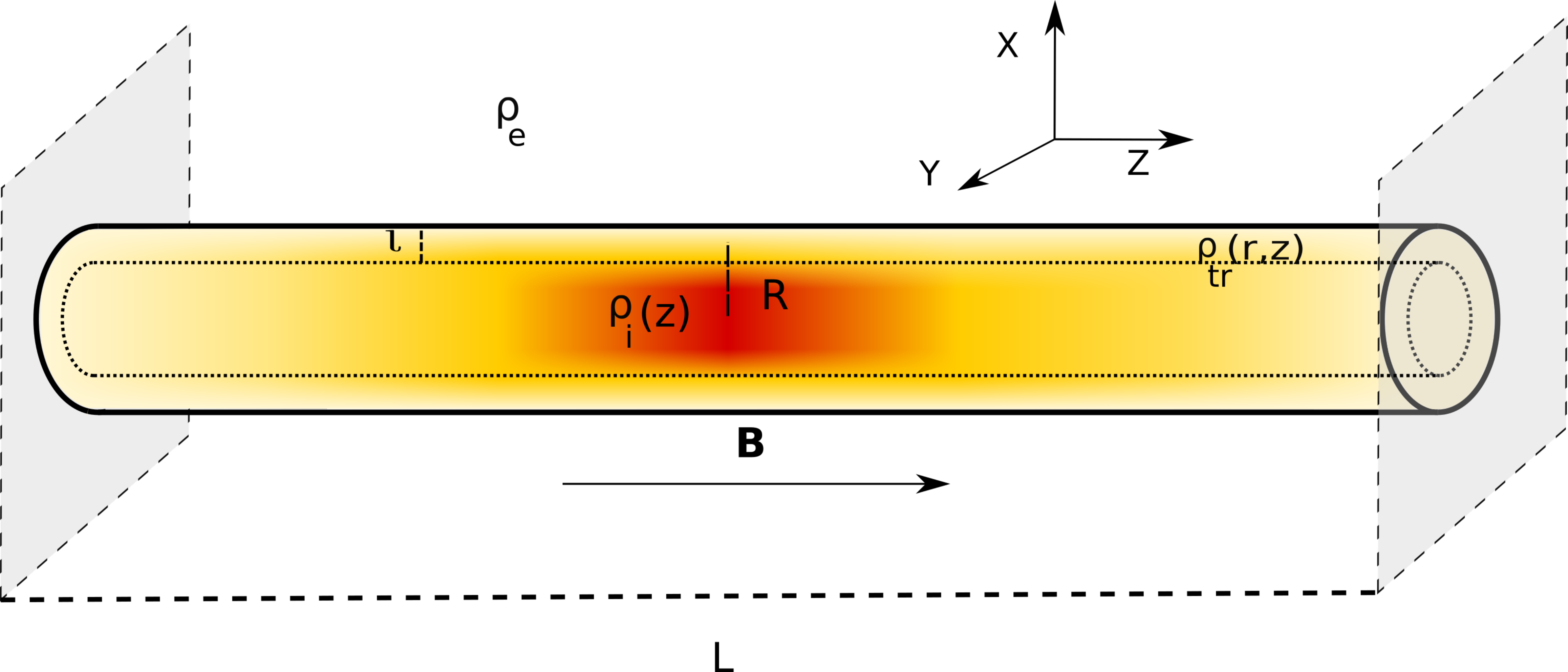}} 
\centering
\caption{Sketch of the prominence thread model. The solar photosphere is represented by the two gray planes at both ends of the magnetic flux tube.}
\label{Fig_model}
\end{figure*}

\begin{figure}
\resizebox{\hsize}{!}{\includegraphics{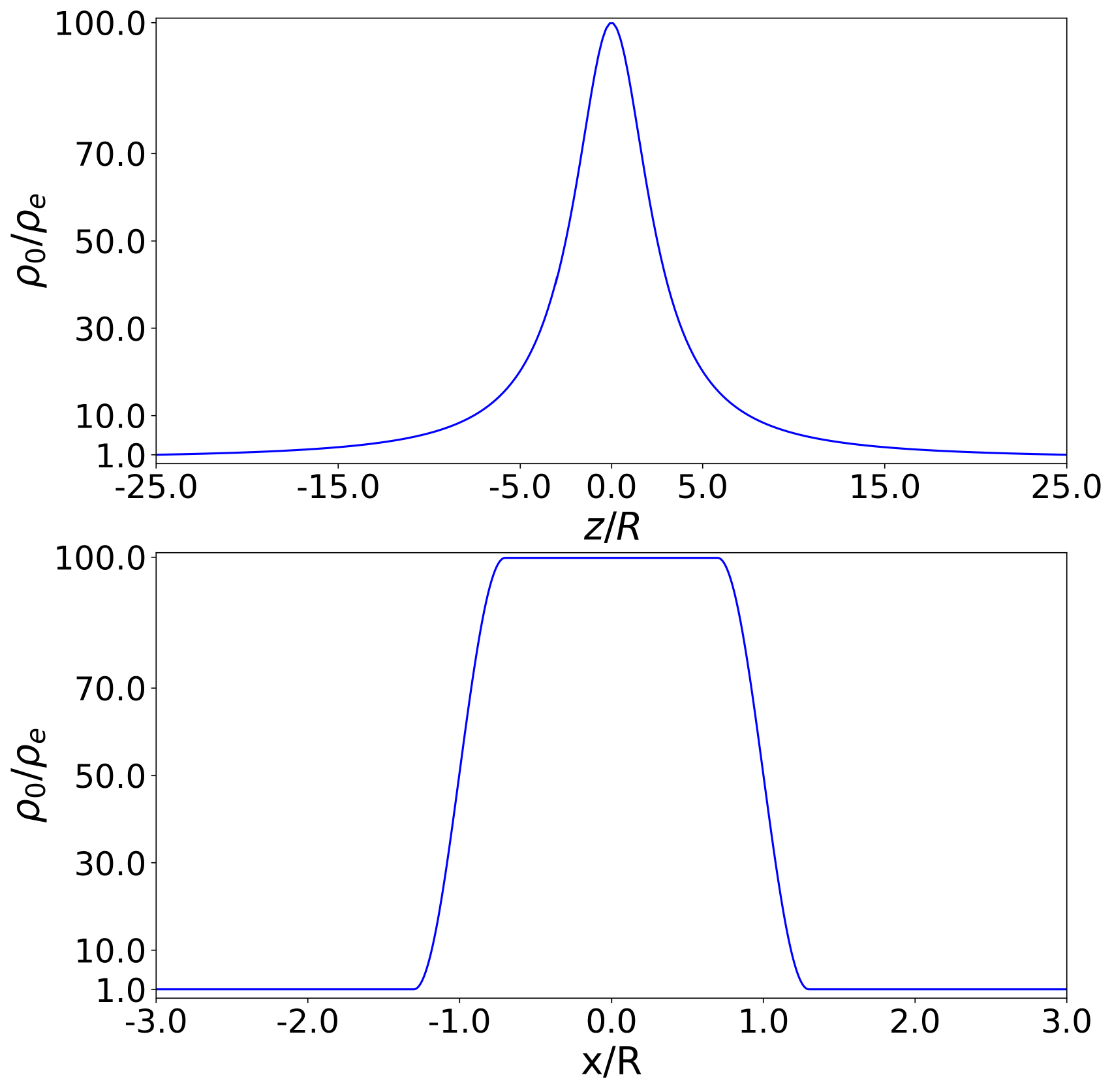}} 
\centering
\caption{Equilibrium density profile. \textit{Top panel:} Longitudinal dependence at $r=0$. \textit{Bottom panel:} Transverse dependence at $y=z=0$. We used $l/R=0.6 $, $\rho_{\mathrm{i,0}}=100\rho_{\mathrm{e}}$, $L/R=50 $, and $ \chi=100 $. The density profiles are normalized with respect to the external, coronal density.}
\label{Fig_rho}
\end{figure}

The cool and dense plasma of the prominence thread is surrounded by the hot and light coronal plasma. For simplicity, we ignore gravity and specify the equilibrium density, $\rho_0$, that varies in both the radial, $r$, and longitudinal, $z$, directions namely, 
\begin{equation}
\rho_{0}(r,z)=
\left\{\begin{array}{lll}
\rho_{\mathrm{i}} (z),  & \mbox{if } & r \leq R-\frac{l}{2}, \\
\rho_{\mathrm{tr}}(r,z),  & \mbox{if } &  R-\frac{l}{2}< \; r  < R+\frac{l}{2}, \\
\rho_{\mathrm{e}},  & \mbox{if } & r \geq R+\frac{l}{2},
\end{array}
\right.
\label{rhoir}
\end{equation}
where $\rho_{\mathrm{i}} (z)$ is the internal density that varies along the tube,  $\rho_{\mathrm{e}} $ is the external coronal density assumed uniform, and $\rho_{\mathrm{tr}}(r,z)$ is the density in a transversely nonuniform layer of thickness, $l$, that connects the internal and external plasmas. In this work we used $l/R = 0.6$. After \citet{soler2015periodratio}, we adopt a Lorentzian profile along the flux tube for the internal density, namely, 
\begin{equation}
\rho_{\mathrm{i}}(z)=\frac{\rho_{\mathrm{i,0}}}{1+4\left(\chi-1\right)z^{2}/L^{2}}.
\label{rhoivert}
\end{equation}
In Eq. (\ref{rhoivert}), $ \rho_\mathrm{i,0} $ is the density at the centre of the flux tube, $z=0 $,  and $ \chi\equiv \rho_{i,0}/\rho_{i}(z=L/2) $ is the  ratio of the central density to the footpoint density. The larger the value of $\chi$, the more concentrated the distribution of the density is around the center of the flux tube \citep[see, e.g., Fig.~2 of][]{Martinez22}. In turn, the density in the  transversely nonuniform layer is prescribed as
\begin{eqnarray}
\rho_{\mathrm{tr}}(r,z)=\frac{\rho_{\mathrm{i}}\left(z\right)}{2}\left\{\left[1+\frac{\rho_\mathrm{e}}{\rho_{\mathrm{i}}\left(z\right)}\right]-\left[1-\frac{\rho_\mathrm{e}}{\rho_{\mathrm{i}}\left(z\right)}\right]\sin\left[\frac{\pi}{l}\left(r-R\right)\right]\right\}.
\label{rhotr}
\end{eqnarray}
In our background model we used  $\rho_{\rm e}=5.02 \cdot 10^{-13}\;\mathrm{kg/m^{3}}$, $\rho_{\rm i,0}=100\rho_{\rm e}=5.02 \cdot 10^{-11}\;\mathrm{kg/m^{3}}$, and $\chi = 100$. Figure~\ref{Fig_model} shows a sketch of the prominence thread model and Fig.~\ref{Fig_rho} shows one-dimensional longitudinal and radial cuts of the equilibrium density profile.

\begin{figure}[htbp!]
\resizebox{\hsize}{!}{\includegraphics{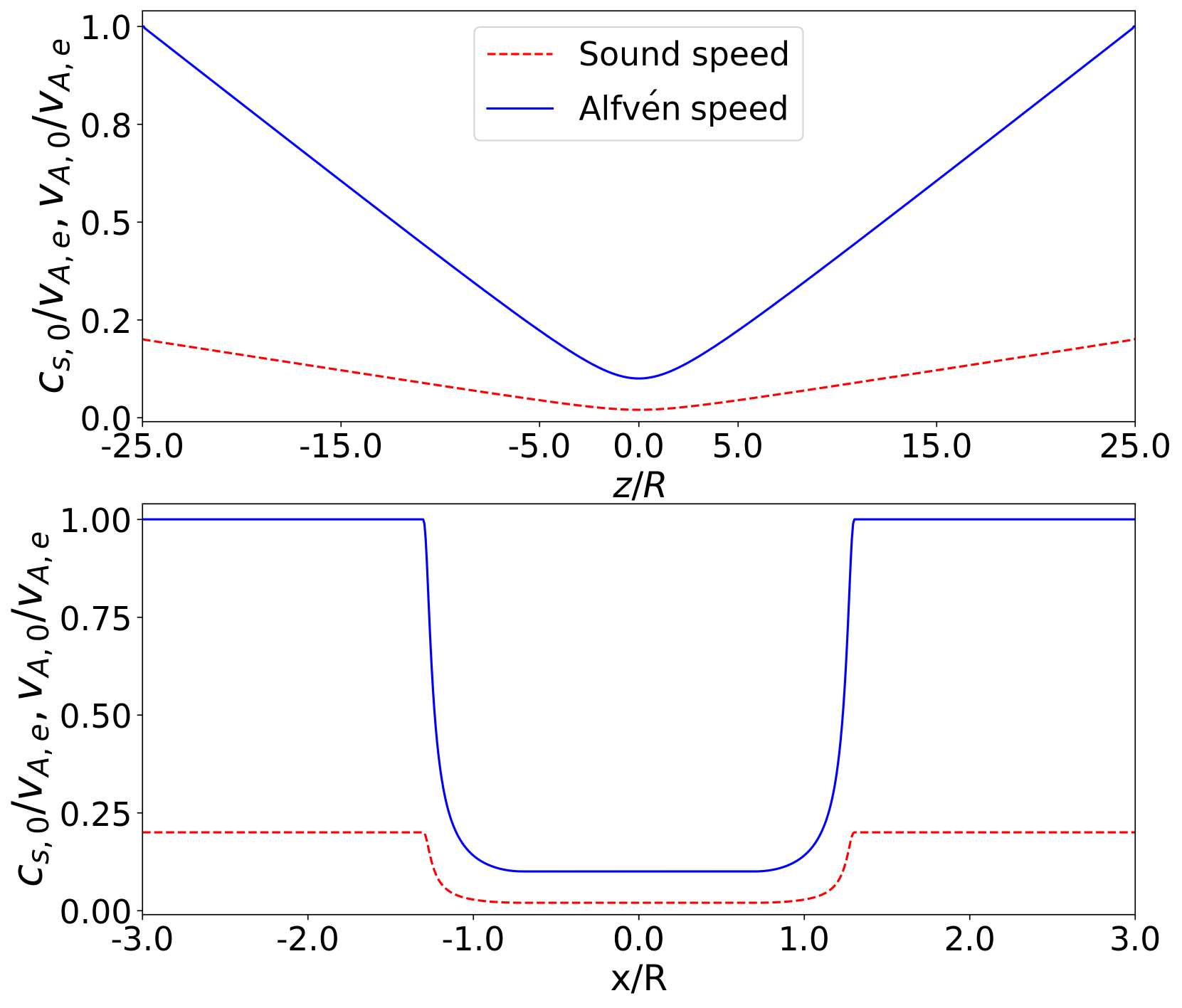}} 
\centering
\caption{Same as Fig.~\ref{Fig_rho}, but for the sound and Alfv\'en speeds in the equilibrium. In both panels, the blue solid line corresponds to the  Alfv\'{e}n speed and the red dashed line corresponds to the sound speed. Both speeds  are normalized with respect to the external Alfv\'{e}n speed, $v_{\rm A,e}$.}
\label{Fig_speeds}
\end{figure}

The background gas pressure in the model, $p_0$, is uniform. In the realistic corona and prominences, the plasma is magnetically dominated, so that the plasma $\beta=2 p_{0}\mu_{0}/B^{2}_{0} \ll 1$, where $\mu_{0} $ is the vacuum magnetic permeability. In our model, we chose the value of $p_0$ so that $\beta = 0.048$. The equilibrium Alfv\'{e}n speed, $v_{\rm A,0} (r,z)=B_{0}/\sqrt{\mu_{0} \rho_{0}(r,z)} $, and the equilibrium sound speed, $ c_{s,0} (r,z)=\sqrt{\gamma p_{0}/ \rho_{0}(r,z) }$, where $ \gamma $ is the adiabatic constant, are displayed in Fig.~\ref{Fig_speeds} in longitudinal and radial cuts to the thread. We can see a sharp variation in  $v_{A,0}$ and $c_{s,0}$ across the prominence thread owing to the large density contrast between the corona and the core of the thread. The variation of both speeds is smoother in the longitudinal direction. The external value of the Alfv\'en speed is our reference velocity, namely $ v_{\rm A,e}=654.7 \;\mathrm{km/s}$.

\subsection{Numerical code}
\label{SUB_NC}

We performed time-dependent numerical simulations with the PLUTO code \citep{Mignone07}, which solves the 3D resistive MHD equations with a finite-volume, shock-capturing spatial discretization. The equations solved by PLUTO are as follows:
\begin{eqnarray}
\frac{\partial\rho}{\partial t} &=& -\nabla \cdot\left(\rho \mathbf{v}\right), \label{cont} \\
\rho \frac{\mathrm{D}\mathbf{v}}{\mathrm{D} t} &=&  -\nabla p + \frac{1}{\mu_0} \left(\nabla \times \mathbf{B}\right) \times \mathbf{B}, \label{mome} \\ 
\frac{\partial \mathbf{B}}{\partial t} &=& \nabla \times \left(\mathbf{v}\times \mathbf{B}\right)-\mu_0 \nabla \times \left(\hat{\eta} \cdot \mathbf{j}\right), \label{indu} \\
\frac{\mathrm{D} p}{\mathrm{D} t} &=& \frac{\gamma p}{\rho}\frac{\mathrm{D} \rho}{\mathrm{D} t}+ \left(\gamma-1\right) \mu_{0}\left(\hat{\eta} \cdot \mathbf{j}\right) \cdot \mathbf{j}. \label{ener} 
\end{eqnarray}
In Eqs. (\ref{cont}-\ref{ener}), $ \frac{\mathrm{D}}{\mathrm{D}t}  \equiv \frac{\partial}{\partial t}+ \mathbf{v} \cdot\nabla$ denotes the total  derivative, $ \rho $ is the mass density, $ \mathbf{v} $ is the velocity,  $ \mathbf{B} $ is the magnetic field, $ p$ is the gas pressure, and $\bf j = \left(\nabla \times {\bf B}\right)/\mu_0$ is the current density. Moreover, $\hat\eta$
 is the resistivity tensor, which is defined in PLUTO as a diagonal tensor in the Cartesian coordinate frame, namely
 \begin{equation}
     \hat{\eta} = \left( \begin{array}{ccc}
     \eta_x & 0 & 0 \\
     0 & \eta_y & 0 \\
     0 & 0 & \eta_z
     \end{array}\right), \label{eq:etatensor}
 \end{equation}
with $\eta_x$, $\eta_y$, and $\eta_z$ the $x$-, $y$-, and $z$-components of resistivity, respectively. 

The PLUTO code solves Eqs.~(\ref{cont}-\ref{ener}) in Cartesian coordinates. We used a uniform grid of 800x800x250 points. We performed the simulations in a computational box where $x/R \in \left[-3,3\right]$, $ y/R \in \left[-3,3\right] $, and $z/R \in \left[-25,25\right] $. Therefore, the spatial resolution in $x-$ and $y-$directions is 7.5~km while in $z-$direction is~200 km.  We used a fifth-order weighted essentially non-oscillatory (WENO) algorithm for spatial reconstruction \citep{Borges08} and  a Roe-Riemann \citep{Roe81} solver to compute the numerical fluxes. Moreover, we used the hyperbolic divergence cleaning technique \citep{Dedner02} to maintain the solenoidal constraint on the magnetic field, which couples the divergence of the magnetic field and Eq. (\ref{indu}) to generalized Lagrange multipliers. Because the magnetic field is current-free and force-free, we used the background-field-splitting technique \citep{Powell94}, which only evolves the magnetic field perturbations. We perform both ideal and resistive MHD simulations. In the ideal simulations ($\hat\eta=0$), an explicit total variation diminishing third-order Runge-Kutta algorithm is used for the temporal evolution. In the resistive simulations ($\hat\eta\neq 0$), the explicit time step becomes too small, as it depends quadratically on the grid size and inversely on the diffusion coefficient. In this case, we use the super-time stepping technique (STS, \citeauthor{Alexiades96} \citeyear{Alexiades96}) implemented in PLUTO. STS allows us to save computational time by accelerating the explicit temporal scheme and requiring stabilization only at the advective time super-steps, but not in the various substeps in which each super-step is divided into.

Although the temperature, $T$, is not a variable directly evolved by the PLUTO code, we are interested in studying its evolution over the span of the simulations. To this end, we implemented in PLUTO the computation of the temperature through an iterative method using the local values of density and gas pressure, so that $T$ is a secondary, user-defined variable of the code. In addition, the prominence plasma is partially ionized \citep[see, e.g.,][]{Parenti14}. Therefore, we also need to determine the plasma ionization fraction, $\xi_{i}$, defined as the ratio of the ion density to the total density. At every time step and at each grid point, we solved the equation of state for a partially ionized hydrogen plasma, namely,
\begin{equation}
p =\left(1+\xi_{i}\right)\rho \tilde{R} T,
\label{eqpip}
\end{equation}
where $\tilde{R}$ is the ideal gas constant. In the prominence plasma, $\xi_{i}$  is a function of the local pressure and temperature. We used the tabulated values of $\xi_{i}$ given in Table~1 of \citet{Heinzel15} for a height of 10~Mm over the photosphere. Equation ~(\ref{eqpip}) is coupled with the values of the table. Starting with the assumption that $\xi_{i} = 1$ (full ionization),  Eq.~(\ref{eqpip}), together with the data in Table~1 of \citet{Heinzel15}, is solved iteratively until the computed values of $T$ and $\xi_{i}$ converge. Since the table only contains limited ranges of pressure and temperature, the following conditions are applied in the process when the values are outside those of the table. Using the results from \citet{Goutterbroze09},  for temperatures higher than $2\cdot 10^{4}$ K, we assumed full ionization. If the pressure is larger or smaller than the maximum or minimum pressure value from the table, or if the temperature is smaller than 6000 K, we saturated to the closest value in the table. However, we note that the condition for the pressure was never actually applied in the simulations included here, as the minimum and maximum pressure values found during the simulations fell within the range of the table in all cases. Particularly, the pressure was found to range between 0.046 and 0.057 dyn~cm$^{-2}$, approximately. Typically, prominences  are composed of $\sim 10 \%$ helium. However, for simplicity, we assumed a plasma  composed solely of hydrogen.

We used an initial perturbation with the aim of exciting the longitudinally fundamental mode of standing torsional Alfv\'{e}n waves. As in \citet{diazsoler21aanda}, we perturbed the azimuthal component of velocity, as follows:
\begin{equation}
\mathbf{v}\left(t=0\right)= v_{0}A(r)\cos\left(\frac{\pi}{L}\;z\right)\; \hat{\varphi},
\label{vazi}
\end{equation}
where $ v_{0} $ is the maximum velocity amplitude and $ A(r)$ contains the radial dependence  \citep[see details in][]{diazsoler21aanda}. A radial cut of the normalized azimuthal component of velocity at the centre of the tube can be seen in top panel of Fig. 4 of \citet{Diazsoler22} (see the red dot-dashed line). We set $ v_{0}= 0.01 v_{A,e}$, so that $ v_{0} = 6.55\; \mathrm{km/s}$, which is of the order of the velocity amplitude in small-amplitude prominence oscillations \citep{Lin09,Arregui18LRSP}. The longitudinal dependence as $\cos\left(\frac{\pi}{L}\;z\right)$, set in the initial condition of Eq.~(\ref{vazi}), excites the fundamental mode as well as other longitudinal harmonics with even symmetry with respect to $z=0$. These additional harmonics will be present because, in a longitudinally nonuniform tube, the spatial dependence of the eigenmodes deviates from the canonical harmonic dependence \citep[see, e.g.,][]{andries2005,soler2015periodratio}. However, in the simulations  the evolution is largely dominated by the dynamics of the longitudinally fundamental mode.

Regarding the boundary conditions, we used outflow conditions, that is, zero gradient for all the variables in all lateral boundaries. On the top and bottom boundaries, $ z=\pm L/2 $, we imposed line-tying conditions so as to mimic the anchoring of the field lines in the solar photosphere. To perform this task, we fixed the three components of velocity to zero while the $z$-component of the magnetic field is fixed to the equilibrium value. The remaining variables, namely density, pressure, and the $x-$ and $y-$components of the magnetic field, are set to be outflow.

\subsection{Implementing Ohmic and ambipolar diffusion}
\label{nonideal}

The temperature of the coronal plasma is typically of the order of $10^{6}$~K \citep[see, e.g.,][]{Priest14}, so the plasma is fully ionized. The temperature in prominences is much smaller, around $10^{4}$~K or lower \citep[see, e.g.,][]{Tandberg95,Parenti14}. As a consequence of that, the prominence plasma is only partially ionized \citep[see, e.g.,][]{Labrosse10}. In a partially ionized plasma, ambipolar diffusion, caused by ion-neutral collisions, is an important dissipation mechanism \citep[see, e.g.,][]{Osterbrock61,Pandey08,Soler09,Khomenko12,Ballester18,Nobrega20}. Similarly, the Ohmic diffusion, caused by the collisions of electrons with other species, is heavily enhanced by the presence of neutrals. Both Ohmic and ambipolar diffusion can dissipate Alfv\'{e}n waves in the partially ionized prominence medium \citep[see, e.g.,][]{Soler09}.

In the presence of Ohmic and ambipolar diffusion, the resistive term $\hat{\eta}\cdot {\bf j}$ in Eqs.~(\ref{indu}) and (\ref{ener}) can be decomposed as \citep[see, e.g.,][]{Bittencourt04},
\begin{equation}
\hat{\eta}\cdot {\bf j}=(\eta_{\rm O}+\eta_{\rm A}\lvert {\bf B} \rvert^{2}) \mathbf{j} -\eta_{\rm A} \left(\mathbf{B}\cdot\mathbf{j}\right)\mathbf{B}, 
\label{etahat}
\end{equation}
where $\eta_{\rm O}$ and $\eta_{\rm A}$ are the Ohmic and ambipolar diffusion coefficients, respectively. The first term on the right-hand side of Eq.~(\ref{etahat}) only includes diagonal elements in the resistivity tensor, $\hat\eta$. This term depends on both $\eta_{\rm O}$ and $\eta_{\rm A}$. Conversely, the second  term introduces both diagonal and off-diagonal elements  and depends on $\eta_{\rm A}$ alone. In the PLUTO code, the resistivity tensor is assumed to be diagonal (Eq.~(\ref{eq:etatensor})). The off-diagonal terms associated with the second term on the right-hand side of Eq.~(\ref{etahat}) cannot be included in the code. Therefore, the implementation of the ambipolar diffusion  in PLUTO  can only be done in an approximate manner \citep[see, e.g.,][for general implementations of the ambipolar diffusion]{Khomenko12,Nobrega20,Moreno22}. To circumvent this limitation of the code, we exploited the fact that, in the simulations,  the background magnetic field along the tube axis is strong compared with the perturbations across the  background magnetic field (see details in the Appendix~\ref{app:approx}). Hence, in  Eq.~(\ref{etahat}) we write
\begin{equation}
\mathbf{B}=B_{0} \hat{z}+\mathbf{B_{1}},
\label{bzper}
\end{equation}
where $ \mathbf{B_{1}} $ represents the perturbations over the background axial field and verify that $\lvert B_{1} \rvert \ll B_{0} $. Consequently, we approximate $\lvert {\bf B} \rvert^{2} = B_0^2$ and $ \left(\mathbf{B}\cdot\mathbf{j}\right)\mathbf{B} \approx B^{2}_{0} {\bf j}\cdot \hat{z} $. Under this approximation,  $\hat\eta$ becomes a diagonal tensor as:

\begin{equation}
 \hat{\eta} \approx \left( \begin{array}{ccc}
     \eta_{\rm O}+B_0^2\eta_{\rm A} & 0 & 0 \\
     0 & \eta_{\rm O}+B_0^2\eta_{\rm A} & 0 \\
     0 & 0 & \eta_{\rm O}
     \end{array}\right) = \left( \begin{array}{ccc}
     \eta_{\rm C} & 0 & 0 \\
     0 & \eta_{\rm C} & 0 \\
     0 & 0 & \eta_{\rm O}
     \end{array}\right) ,
\label{etadiagonal}
\end{equation}
where $\eta_{\rm C} = \eta_{\rm O}+B_0^2\eta_{\rm A}$ is the Cowling's diffusion coefficient, which combines both the ambipolar and Ohmic diffusion coefficients. The Cowling diffusion accounts for the total magnetic diffusion across the magnetic field. We implemented in the PLUTO code this approximate resistivity tensor. Some tests about the numerical implementation  can be checked in the Appendix~\ref{AppendixB}.

The Ohmic, $\eta_{\rm O}$, and ambipolar, $\eta_{\rm A}$, diffusion coefficients depend upon the plasma local properties and are calculated in PLUTO  in the whole domain at each time step. The expressions to compute both coefficients are \citep[see, e.g.,][]{Ballester18}:
\begin{equation}
\eta_{\rm O}= \frac{\alpha_{e}}{\mu_{0}e^{2}n^{2}_{e}},
\label{etaohm}
\end{equation}
\begin{equation}
\eta_{A}= \frac{\xi^{2}_{n}}{\mu_{0}\alpha_{n}},
\label{etaambi}
\end{equation}
where $ \xi_{n} = 1-\xi_{i}$ is the neutral fraction, $e$ is the electron charge, and $n_e$ is the electron number density. Moreover, $\alpha_{e} $ and  $ \alpha_{n} $ are the total friction coefficients for electrons and neutrals, respectively, which account for the collisions that these species have with all other particles. In our particular case, the species can be electrons, protons, or hydrogen atoms because we used a pure hydrogen plasma.  The total friction coefficient of a species $ \beta $ can be obtained as the sum of the individual friction coefficients between different species, namely,
\begin{equation}
\alpha_{\beta}=\sum_{\beta\neq \beta'} \alpha_{\beta \beta'}.
\label{alphane}
\end{equation}
Following \citet{Braginskii65} and \citet{Spitzer68}, the friction coefficient between two different charged species is:
\begin{equation}
\alpha_{\beta \beta'}=\frac{n_{\beta}n_{\beta'}e^{4}\ln{\Lambda_{\beta\beta'}}}{6\sqrt{2}\epsilon^{2}_{0}m_{\beta\beta'}\left(\pi k_{B}T/m_{\beta\beta'}\right)^{3/2}},
\label{alphaei}
\end{equation}
where  $ n_{\beta} $ ($ n_{\beta'})$ is the number density of the species $\beta\; (\beta')$, $m_{\beta\beta'}$ is the reduced mass defined as $ m_{\beta}m_{\beta'}/(m_{\beta}+m_{\beta'})$, $k_B$ is the Boltzmann's constant, and $ \epsilon_{0}$ is the vacuum electrical permittivity. Moreover, $ \ln{\Lambda_{\beta\beta'}}$ is the Coulomb logarithm, which is included to consider the ineffectiveness of the interactions between the charged species after some distance. According to \citeauthor{Spitzer62} (\citeyear{Spitzer62}; see also \citeauthor{Vranjes13} \citeyear{Vranjes13}), the Coulomb logarithm is:
\begin{equation}
\ln{\Lambda_{\beta\beta'}}=\ln\left[\frac{24\pi\left(\epsilon_{0}k_{B}T\right)^{3/2}}{e^{3}\sqrt{n_{\beta}+n_{\beta'}}}\right].
\label{lncoulomb}
\end{equation}
On the other hand, the friction coefficient between a charged and a neutral species is:
\begin{equation}
\alpha_{\beta n}=n_{\beta}n_{n}\sqrt{\frac{8k_{B}T}{\pi m_{\beta n}}} \sigma_{\beta n},
\label{alphabetan}
\end{equation}
where the sub-index $n$ denotes a neutral and $ \sigma_{\beta n}$ is the collisional cross-section, which we obtained from \citet{Vranjes13}. The values of the collisional cross-section are weakly dependent on the temperature \citep[see][]{Vranjes13}. However, for typical prominence temperatures considered here, this result can be safely ignored. Thus, we used constant values of the collisional cross-sections for simplicity.

\begin{figure*}
\includegraphics[width=17cm]{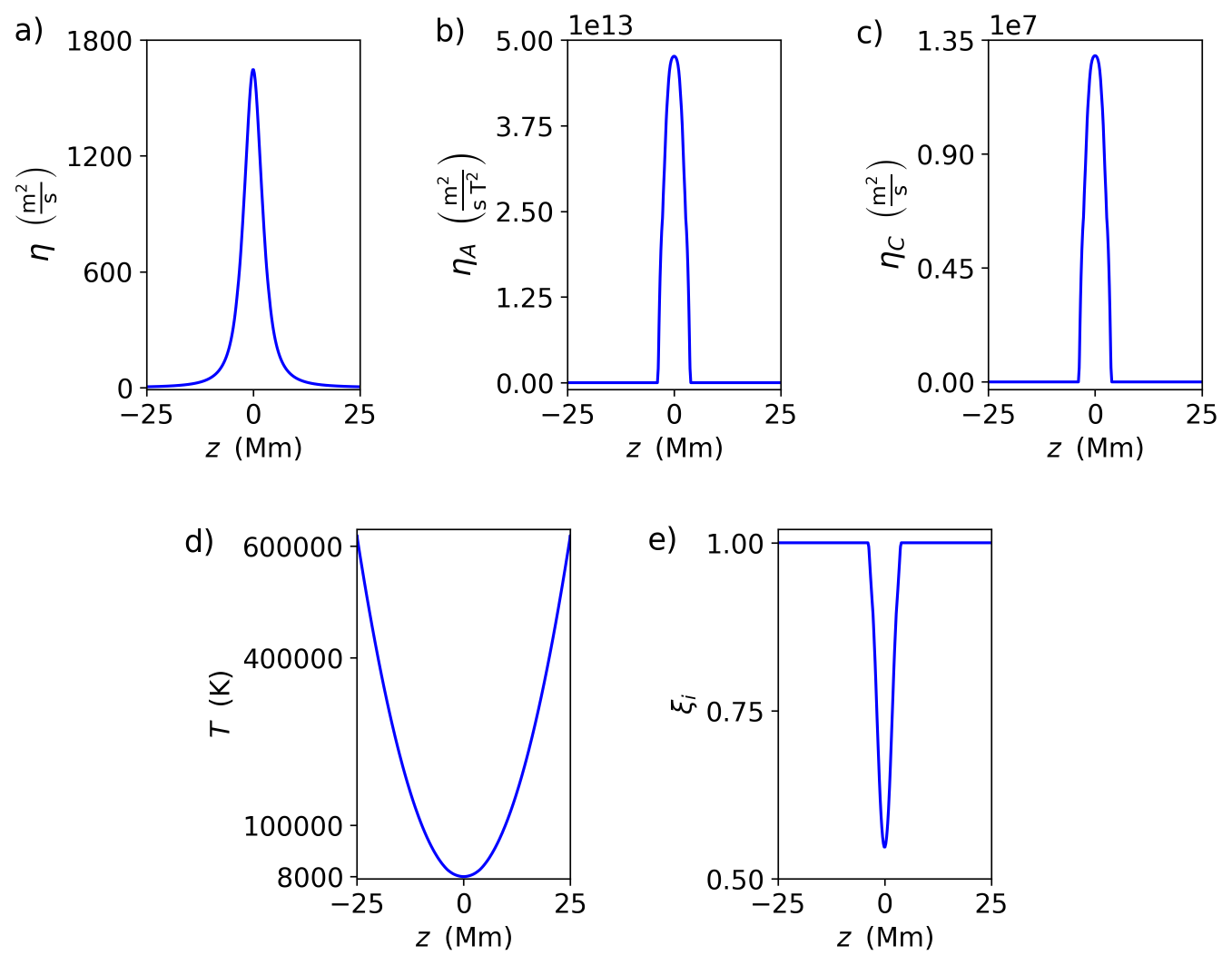} 
\centering
\caption{Longitudinal cuts along the tube axis, $r=0$, of  the \textit{a)}  Ohmic diffusion coefficient, \textit{b)} ambipolar diffusion coefficient, \textit{c)} Cowling diffusion coefficient, \textit{d)}  temperature, and \textit{e)} ionization fraction in the equilibrium.}
\label{Fig_etastempxiinitial}
\end{figure*}

We studied the values of the temperature, the ionization fraction, and the Ohmic, ambipolar, and Cowling diffusion coefficients in a longitudinal cut at the tube axis, $r=0$, corresponding to the background model. The results are shown in Fig. \ref{Fig_etastempxiinitial}. We found the minimum of temperature, $T \approx 8000 $~K, and the minimum of the ionization ratio, $\xi_{\rm i}\approx 0.547 $, at the core of the prominence thread.  The ambipolar diffusion is dominant for typical conditions of prominence threads, so it is the main contributor to the Cowling diffusion \citep[see][]{Melis21,Melis23}. A similar situation occurs in the high chromosphere \citep{Khomenko14phpl}. Magnetic diffusion is highly anisotropic in the partially ionized prominence plasma, being much more efficient across than along the magnetic field direction.

\section{Numerical simulations}
\label{sec:simulations}

We aim to study the nonlinear evolution of standing torsional oscillations in the prominence thread. Unless otherwise stated, all quantities are specified in normalized units. Density, velocity, and length are normalized with respect to the external density, $ \rho_{e} $, external Alfv\'en speed, $ v_{\rm A,e}$, and the thread radius, $R$, respectively, whose values have been given before. In turn, the normalized time, $\bar t$, is expressed in units of the Alfv\'{e}n travel time, namely, $ t_{\rm A}=R/v_{\rm A,e}\approx 1.53 $~s.

Before analyzing the result of the simulations, it is useful to calculate the expected period of the standing oscillations. Following the analytical treatment in \citet{diazsoler21aanda}, to calculate the period of the longitudinally fundamental torsional Alfv\'{e}n mode we solved the 1D wave equation for the linear azimuthal component of velocity, $ v_{\varphi}' $, in the equilibrium, namely,
\begin{equation}
\frac{\partial^{2}v_{\varphi}'}{\partial t^{2}}=v^{2}_{\rm A,0}(r,z)\frac{\partial^{2}v_{\varphi}'}{\partial z^{2}}. 
\label{torsional}
\end{equation}
Equation (\ref{torsional}) is transformed into an eigenvalue problem by expressing the temporal dependence as $ \exp \left(  i\omega_{\rm A} t\right)$ where $\omega_{\rm A}  $ is the Alfv\'en eigenfrequency.  Thus,
\begin{equation}
\omega^{2}_{\rm A} v_{\varphi}'= -v^{2}_{\rm A,0}(r,z)\frac{\partial^{2}v_{\varphi}'}{\partial z^{2}},
\label{torsionaleigen}
\end{equation}
with the boundary conditions of $ v_{\varphi}'(r,z =\pm L/2)=0 $. Because of the spatial dependence of $v_{\rm A,0}$, we solved the eigenvalue problem numerically with Wolfram Mathematica for the longitudinally fundamental mode. The period of the oscillations is $ P= 2\pi/\omega_{\rm A} $. The period varies with the radial position because of the radial dependence of $v_{\rm A,0}$. The period in the uniform core of the thread  is $\sim 535\,t_{\rm A}$ (13.62 min in physical time), while that in the corona  is $100\,t_{\rm A}$ (2.54 min in physical time). This means that the Alfv\'en modes in the external plasma would have completed more than 5 periods of oscillation before the internal mode had completed a single period. Of course, the period continuously varies in the transition between the internal and external plasmas. The result of this continuum of periods will be a strong phase mixing in the radial direction.

\subsection{Ideal MHD dynamics}
\label{Sub_idealdyn}

\begin{figure*}
\includegraphics[width=17cm]{{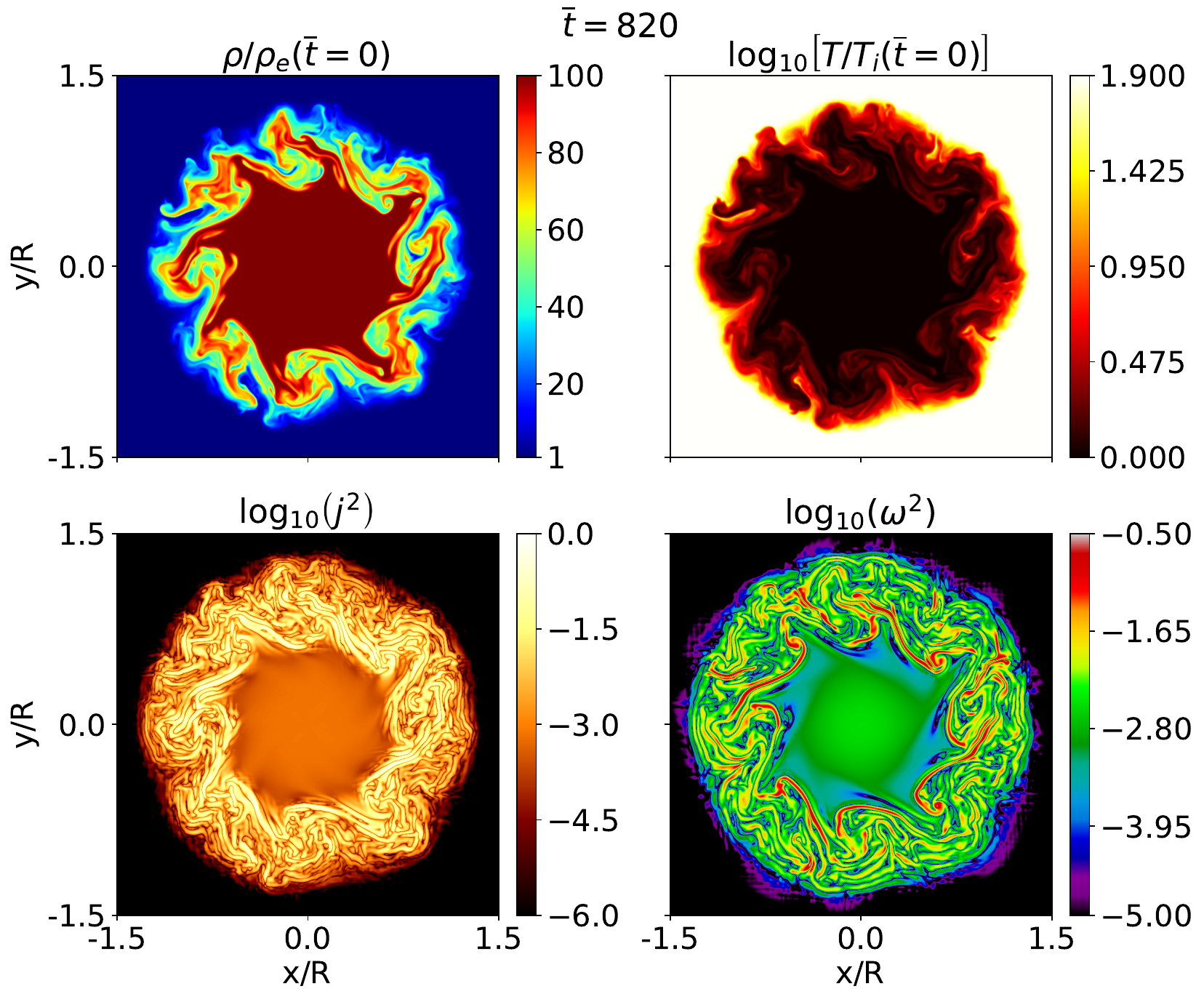}} 
\centering
\caption{Cross-sectional cuts  of density (\textit{top left}), temperature (\textit{top right}), current density squared (\textit{bottom left}), and vorticity squared (\textit{bottom right}) at the end of the ideal MHD simulation, $\overline{t}=820 $. Temperature is normalized with respect to the initial value at  $r=0$. Logarithmic scale is used for each variable except for the density to optimize visualization. The cross-sectional cuts are done at the tube center, $z=0$, except for current density squared which is done at $z=R$.  The complete temporal evolution is available  as an online movie.}
\label{Fig_zoplanes}
\end{figure*}

\begin{figure*}[htbp!]
\includegraphics[width=17cm]{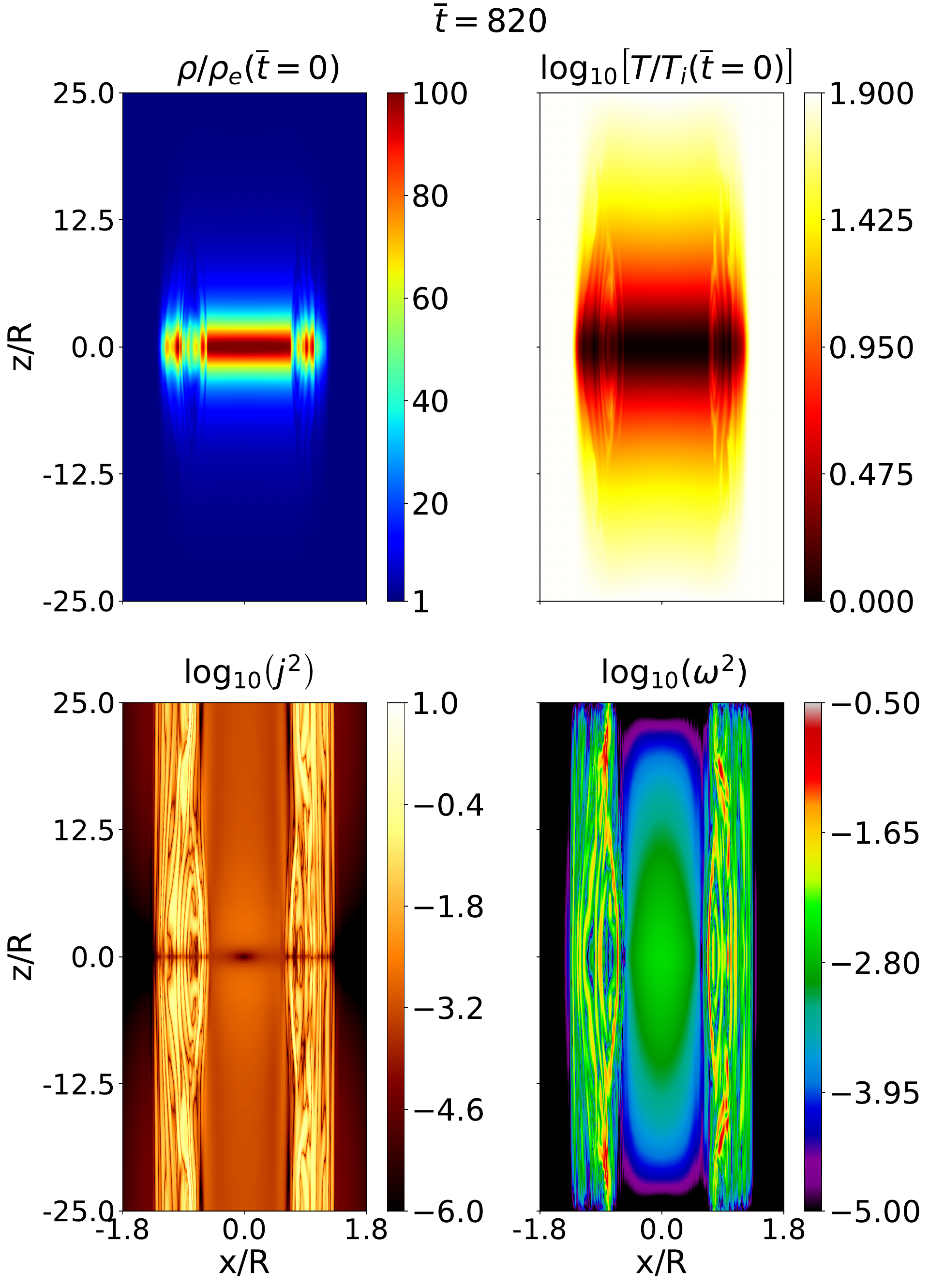} 
\centering
\caption{Same as Fig. \ref{Fig_zoplanes}, but the cut is done longitudinally at $ y=0$. Note: the horizontal and vertical axes are not to scale. The complete temporal evolution is available  as an online movie.}
\label{Fig_yoplanes}
\end{figure*}

First, we studied the dynamics of the oscillations in the absence of resistivity, so that we set $\eta_{\rm O}=\eta_{\rm A} =0$ and performed an ideal MHD simulation. After the initial excitation, the flux tube is let to evolve. To help us understand better the dynamics of the simulation, besides studying the evolution of the density and the temperature, we also studied the vorticity, $ \boldmath{\omega}= \nabla \times \mathbf{v}$, and the current density, ${\bf j}=\mu_{0}^{-1}\nabla \times \mathbf{B}$ because these quantities are extremely sensitive to spatial gradients in the velocity field and in the magnetic field, respectively. They can help us visualize the development of turbulence. The results of this ideal simulation are displayed in Figs.~\ref{Fig_zoplanes} and \ref{Fig_yoplanes} in two different planes of the complete 3D box.

On the one hand, Fig.~\ref{Fig_zoplanes} shows cross-sectional cuts of the density (top left panel), the temperature (top right panel), the current density squared (bottom left panel), and the vorticity squared (bottom right panel). We used logarithmic scale to optimize the visualization for each variable except for the density.  The cross-sectional cuts for density, vorticity squared, and temperature are done at the tube center, $z=0$, while that for current density squared  is done at $z=R$, which is a plane slightly displaced from the center. The reason for showing the current density in a different plane is that this variable has a node at the tube center. Since the relevant dynamics occurs in the core and in the transition layer of the prominence thread, we only showed a subdomain of the computational box where $ x,y \in [-1.5R, 1.5R]$. We checked that nothing of interest happens outside this subdomain. The complete temporal evolution can be seen in the accompanying animation, while the still image in Fig.~\ref{Fig_zoplanes} displays the results at the final simulation time, $\overline{t}=820 $, corresponding to $\sim 21$ min in physical time.

On the other hand, Fig.~\ref{Fig_yoplanes} shows the same variables as in  Fig.~\ref{Fig_zoplanes}, but in a longitudinal plane to the tube corresponding to $y=0$. As before, the results are only displayed in a subdomain of the whole box where $x \in [-1.8R, 1.8R]$, while the $z-$direction is fully shown. In Fig.~\ref{Fig_yoplanes} we used a larger scale for the current density than in Fig.~\ref{Fig_zoplanes} to avoid early saturation and ease the visualization. Again, the complete temporal evolution can be seen in the accompanying animation, while the still image in Fig.~\ref{Fig_yoplanes} displays the results at the final simulation time.

As expected, we found that the torsional Alfv\'{e}n modes oscillate with different frequencies in adjacent radial positions owing to the transversely nonuniform density. The oscillations become out of phase as time increases. Such a phenomenon is called phase mixing and is of linear nature \citep[see, e.g.,][]{HeyvaertPriest83,Nocera84,Moortel00,Smith07,Prokopyszyn2019,Ebrahimi2020,Vandamme20,Ebrahimi2022}. Due to phase mixing, the azimuthal component of velocity (not shown here) alternates between negative and positive values in adjacent radial positions. Such alternation generates azimuthal shear flows (see, e.g., Fig. 2 in \citeauthor{HeyvaertPriest83} \citeyear{HeyvaertPriest83} or Fig. 6 in \citeauthor{diazsoler21aanda} \citeyear{diazsoler21aanda}). These shear flows manifest themselves in the cross-sectional cuts of the vorticity and the current density as concentric rings, which are better seen in the animation of Fig. \ref{Fig_zoplanes}). Similar ring-shaped structures can also be found  in simulations of torsional oscillations of coronal loops \citep{diazsoler21aanda,Diazsoler22}. These early structures in vorticity and current density are also equivalent to those seen in simulations of kink oscillations of coronal loops \citep{Antolin14,Howson17a,Antolin19}. However, the spatial distribution in the case of kink oscillations is not in the form of concentric rings owing to the different azimuthal symmetry that torsional and kink modes have. Besides, resonant absorption  does not happen in our simulation, which is a concurrent process to  phase-mixing  in simulations of kink modes.

The annular structures  in vorticity and current density caused by phase mixing are seen in the simulations before any appreciable disturbance in the density or the temperature. At the early stages of the evolution, we can see some weak perturbations in density in the form of periodic accretion (voiding) of plasma to (from) the center of the thread, which is better appreciated in the animation of Fig.~\ref{Fig_yoplanes}. These density variations are not related to phase mixing but are caused by the ponderomotive force, a nonlinear effect that couples Alfv\'en and slow magnetosonic modes \citep[][]{Hollweg71,Rankin94,Tikhonchuk95,Terradas04}.

As phase mixing develops, the azimuthal shear flows gradually intensify and eventually trigger the KHI \cite[see, e.g.,][]{HeyvaertPriest83,Browning84,soler2010,Zaqarashvili2015,Barbulescu19apj}. As these flows are perpendicular to the background magnetic field lines, the magnetic tension cannot avoid the triggering of KHI \citep[see, e.g.,][]{Chandrasekhar61book}.  We can visually identify in Fig.~\ref{Fig_zoplanes} that the first KHI-associated deformations of temperature and density occur at \textbf{ $ \tau_{vis} \approx 280 $ } ($\sim 7$~min in physical time), which corresponds to less than three periods of the external torsional Alfv\'{e}n mode and about half a period of the internal mode. However, the examination of the form of the vorticity for that time already reveals the existence of significant deformations  in the boundary between the inhomogenous layer and the external medium. This makes us wonder whether the visual estimation of the KHI onset time from the density evolution might overestimate the actual onset time. Indeed, the initial growth of KHI perturbations can be appreciated in the vorticity at an earlier time than in the density and the temperature. A more detailed analysis of onset time is given in Sect.~\ref{sec:KHi}.

Once the KHI is triggered, large eddies are formed and rapidly grow inside the nonuniform boundary layer of the tube. The KHI evolves nonlinearly, breaking the initially large eddies into small eddies leading naturally to turbulence. Turbulence mixes the hot and light coronal plasma with the cold and dense prominence thread plasma. Initially, turbulence occurs locally in the nonuniform boundary layer alone, but as time increases, part of the core and the external media are also affected.  In the temperature evolution, we can see intrusions of hot plasma towards the cool core of the thread  and, in density, an apparent breakup of the prominence material.  Importantly, neither the KHI nor turbulence are seen in the longitudinal direction (see Fig.~\ref{Fig_yoplanes}), meaning that the turbulence only develops perpendicularly to the magnetic field lines. In the case of torsional oscillations of coronal loops, this matter has been investigated in \citet{diazsoler21aanda,Diazsoler22}.

\subsubsection{Estimating the Kelvin-Helmholtz instability onset time}
\label{sec:KHi}
\begin{figure}[htbp!]
\resizebox{\hsize}{!}{\includegraphics{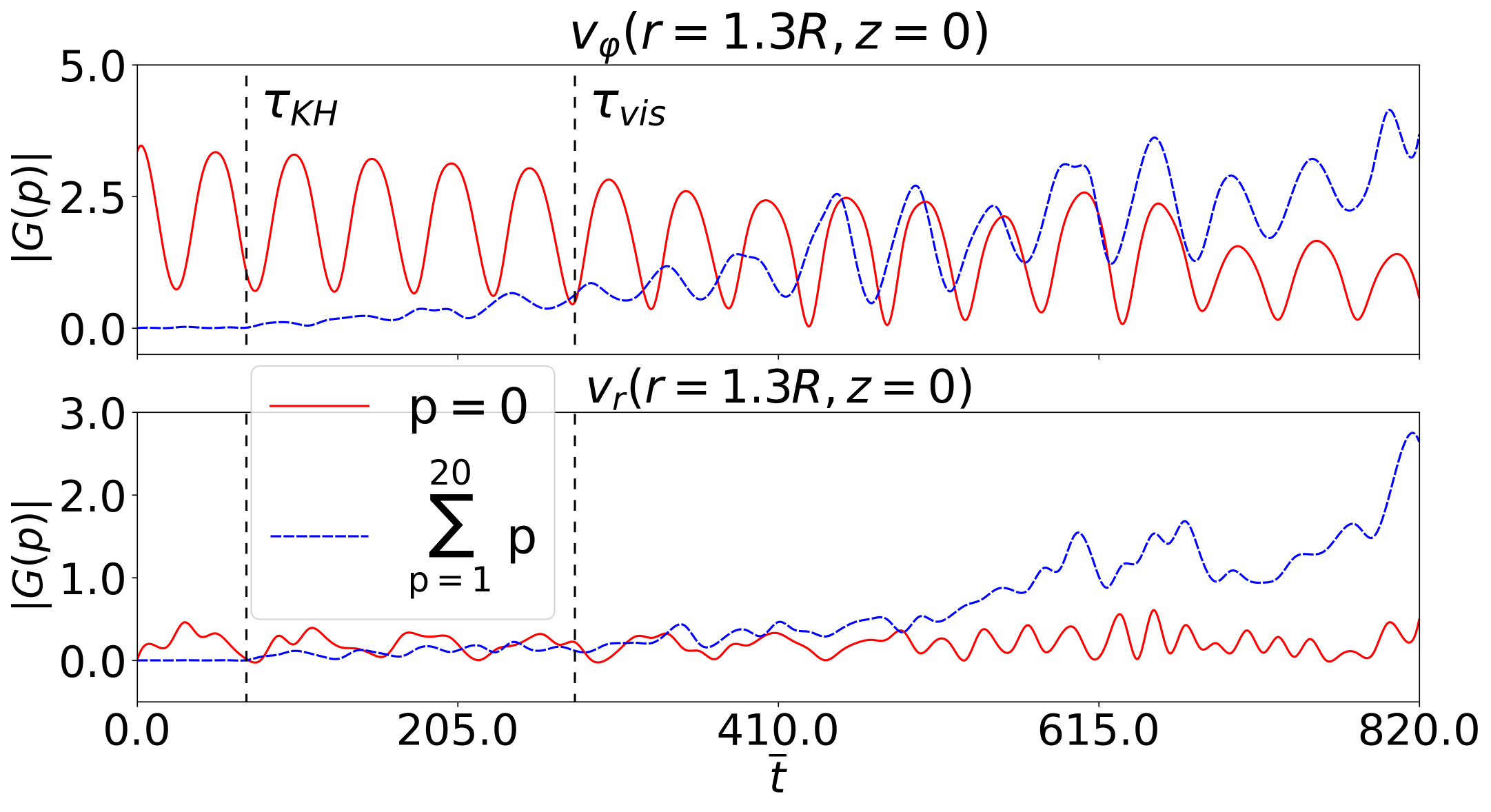}} 
\centering
\caption{\textit{Top panel}: Temporal evolution of the Fourier coefficient with $p=0$ (red solid line) and of the sum of the modulus of the first twenty positive Fourier coefficients (blue  dashed line). For this analysis, the azimuthal component of velocity along a circle of radius $1.3R $ at the $z=0$ plane is used. The two vertical dashed black lines show, respectively, the onset of the KHI obtained from this Fourier analysis, $\tau_{KH} \approx 70$, and the time at which the KHI is first seen to grow in the density, $\tau_{vis}=280$. \textit{Bottom panel}: Same as the top panel but for the radial component of velocity.}
\label{Fig_azimutal}
\end{figure}

To estimate quantitatively the KHI onset time, we used the discrete Fourier transform in the azimuthal direction of the azimuthal and radial velocity components. The KHI is expected to  excite high azimuthal modes. This technique was first used by \citet{Terradas18} and later by \citet{Antolin19} and \citeauthor{diazsoler21aanda} (\citeyear{diazsoler21aanda}, \citeyear{Diazsoler22}) in coronal loops. Particularly, we investigated the azimuthal and radial components of velocity in a cross sectional cut at the tube center, $z=0$, and in the boundary between the external medium and the inhomogeneous density layer, that is, for $r=1.3R$. This radial position is chosen because  we visually identify the growth of the first KHI eddies at that location. We applied the discrete Fourier transform to both velocity components using the fast Fourier transform (FFT) algorithm with the Scipy module \citep{Virtanen20}. Following the same notation as in \citet{Terradas18}, the discrete Fourier transform can be set as:
\begin{equation}
G(p)=\sum^{N-1}_{k=0} g(k) \exp{\left(-\frac{2\pi ipk}{N}\right)}.
\label{fftdiscraz}
\end{equation}
In Eq. (\ref{fftdiscraz}),  $N$ is the number of samples, $g(k)$ is the angular sampling of the azimuthal (radial) velocity, and $p $ is an integer that plays the role of the azimuthal wavenumber and ranges between 0 and $N-1$. In this notation, $p=0$ is the torsional or sausage mode, $p=1$ is the kink mode, and $ p \ge 2$ are the fluting modes \citep[see, e.g.,][]{Roberts19}. Generally, the Fourier amplitudes of Eq.~(\ref{fftdiscraz}) are complex with the exception of $ G(p=0)$. Consequently, we calculated the modulus of all the Fourier coefficients. The temporal evolution of the Fourier amplitudes is shown in Figure \ref{Fig_azimutal}. To optimize the visualization of the contribution of the modes different from $p=0$, which would be the dominant one, we added the first twenty modes starting from $p=1$.

As expected, the $p=0$ mode is dominant during most of the simulation because this is the azimuthal symmetry imposed by the initial excitation. The sum of the first twenty Fourier modes other than $p=0$ is initially zero, as expected. The amplitude of the $p=0$ mode oscillates with a periodicity of 100 in code units, which consistently matches the period of the local torsional Alfv\'en mode at the chosen radial position. As the linear development of phase mixing does not excite higher azimuthal modes, the time at which $ |G (p>0) | $ departs from zero can be defined as the KHI onset time, $\tau_{KH}$. In particular, we find that $\tau_{KH} \approx 70$ in code units. We recall that the periods of the internal and external torsional modes are 535 and 100 in code units, respectively, which indicate that the KHI is very quickly driven in the system. The KHI onset happens significantly earlier than the visually determined time of $ \tau_{vis} \approx 280 $ that is based on the growth of the KHI vortices in density. In fact,  the sum of the first twenty Fourier modes is already comparable with the amplitude of the torsional mode when the KHI eddies become visible in density and in temperature.

A similar analysis but with the inclusion of higher azimuthal modes does not change the obtained results  (not shown here). However, when the analysis is done in layers of the transition region nearer the core, the period of the torsional mode increases, which is expected, and the growth of the other Fourier modes is delayed. The chosen radial position of $r=1.3R$ corresponds to the location where the KHI first grow, so that the analysis at that location provides the smallest KHI onset time.

\subsubsection{Evolution of integrated vorticity and current density}
\begin{figure}[htbp!]
\resizebox{\hsize}{!}{\includegraphics{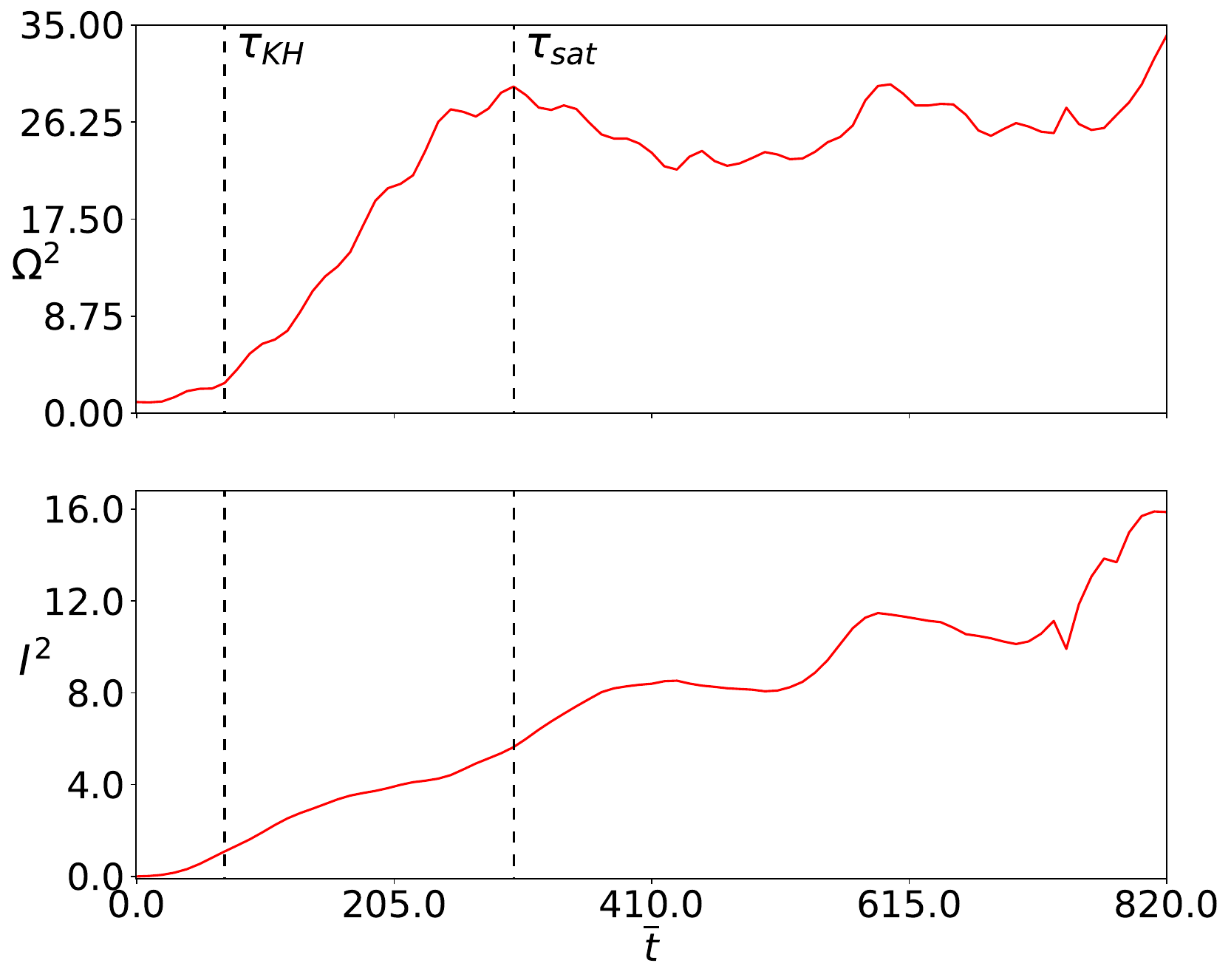}} 
\centering
\caption{Temporal evolution of the vorticity squared (\textit{top panel}) and the current density squared (\textit{bottom panel}) integrated in the whole computational domain. The vorticity  values are normalized with respect to the initial value. The two dashed black lines show respectively the onset of the Kelvin-Helmholtz instability inferred from the analysis of azimuthal modes, $\tau_{KH}=70, $ and the time at which the integrated vorticity squared saturates, $\tau_{sat}=300$.}
\label{Fig_vortycurry}
\end{figure}

In the later stages of the simulation when turbulence plays a predominant role, we  notice the very small scales that are generated in the current density and the vorticity. In addition, the values of the two quantities seem to increase with time during both the initial quasi-linear stage governed by phase mixing and the later nonlinear stage governed by turbulence. To quantify the increase,  we calculated the vorticity squared and the current density squared integrated in the whole computational domain as functions of time, namely
\begin{equation}
\Omega^{2} (t)=\int \left|\boldmath{\omega} (\mathbf{r},t)\right|^{2} \mathrm{d}^3 \mathbf{r},
\label{omega}
\end{equation}
\begin{equation}
I^{2} (t)=\int \left|{\bf j} (\mathbf{r},t)\right|^{2} \mathrm{d}^3 \mathbf{r}.
\label{curr}
\end{equation}
The calculations are shown in Fig.~\ref{Fig_vortycurry} after normalizing $\Omega^{2}$ with respect to the initial value. The background model is current-free, so that $I^2 = 0$ at $t=0$. Using the vorticity evolution in the top panel of Fig. ~\ref{Fig_vortycurry}, we can distinguish three stages in our simulation: a quasi-linear phase, a nonlinear phase dominated by the KHI growth, and a saturation phase where turbulence develops. The two vertical black dashed  lines in Fig.~\ref{Fig_vortycurry} denote the transitions between these distinct stages.

In the short-lived quasi-linear phase, there is an approximately linear increase of $\Omega^{2}$ with time  owing to the slow building up of small scales by phase mixing \citep{Soler15,Howson2020,diazsoler21aanda}. This stage ends when the KHI is triggered for $t=\tau_{KH}$, which coincides with a change of trend in the $\Omega^{2}$ curve. This coincidence reinforces the validity of the Fourier analysis to determine the KHI onset time. During the second phase, the KHI development more rapidly increases the values of vorticity \citep[see][]{Howson17a,Guo19,diazsoler21aanda,Diazsoler22}. This large increase in vorticity should  continue indefinitely in our ideal simulation as the large vortices break into smaller vortices during the KHI turbulent evolution. Nevertheless, the increase in $\Omega^2 $ stops and saturates at $t=\tau_{sat}$, with $\tau_{sat} \approx$~300 in code units for this particular simulation. The saturated phase is characterized by the development of turbulence in the flux tube. The limited spatial resolution and its associated numerical diffusion is behind this saturation. Having a high spatial resolution is crucial to  capture the finest structures generated by the  turbulence \citep{Howson17bres,diazsoler21aanda}.

There is also an increase in the values of $I^2$, but much smoother than that observed for $\Omega^{2}$. The three separate phases discussed before are not so clearly distinguishable in the evolution of $I^2$. One may ask why both variables have different behaviors. To answer that question, one needs to consider the spatial dependence of vorticity and current density along the flux tube. For a standing torsional Alfv\'{e}n wave, the velocity perturbations have nodes at the tube footpoints and an antinode at the tube apex. The vorticity follows the same dependence. Therefore, in Eq.~(\ref{omega}) the largest contribution to the integral of $\Omega^{2}$ comes from the apex of the tube. In turn, the magnetic field perturbations have antinodes at the tube footpoints and a node at the tube apex, with the current density mimicking the same behavior. Hence, in Eq. (\ref{curr}) the largest contribution to the integral of $I^{2}$  originate from the footpoints of the magnetic flux tube. Since the KHI and turbulence develop predominantly  around the tube apex, they heavily impact on the evolution of $\Omega^{2}$. Conversely, the KHI and turbulence do not have such a pronounced development at the tube footpoints, where phase-mixing remains as the dominant mechanism that creates small scales over time. As a result, there is a less significant imprint of the KHI and turbulence in the evolution of $I^{2}$, which keeps displaying the linear growth caused by phase mixing even after $\Omega^{2}$ has already saturated.


\subsection{Effect of Ohmic and ambipolar diffusion}
\label{Subsect_res}

The generation of small structures in the current density and the increase of its value over time might lead to the dissipation of the wave energy  if magnetic diffusion is included. Here we focus on studying the role of Ohmic and ambipolar diffusion. To this end, we ran two additional simulations. These new simulations were performed under the same conditions as the ideal simulation but adding the resistive term in the equations. In the first simulation, called the Ohmic simulation,  we included Ohmic diffusion, but not ambipolar diffusion. Consequently, no approximation is used in the treatment of resistivity and the PLUTO resistivity tensor is in this case isotropic, namely
\begin{equation}
 \hat{\eta} = \left( \begin{array}{ccc}
     \eta_{\rm O} & 0 & 0 \\
     0 & \eta_{\rm O} & 0 \\
     0 & 0 & \eta_{\rm O}
     \end{array}\right),
\end{equation}
In the second simulation, called the Cowling simulation,  we included both ambipolar diffusion and Ohmic diffusion, so the resistivity tensor is anisotropic and follows Eq.~(\ref{etadiagonal}). The purpose in this Subsect. is to compare the two dissipative simulations with the previously discussed ideal MHD simulation.

\begin{figure*}
\includegraphics[width=17cm]{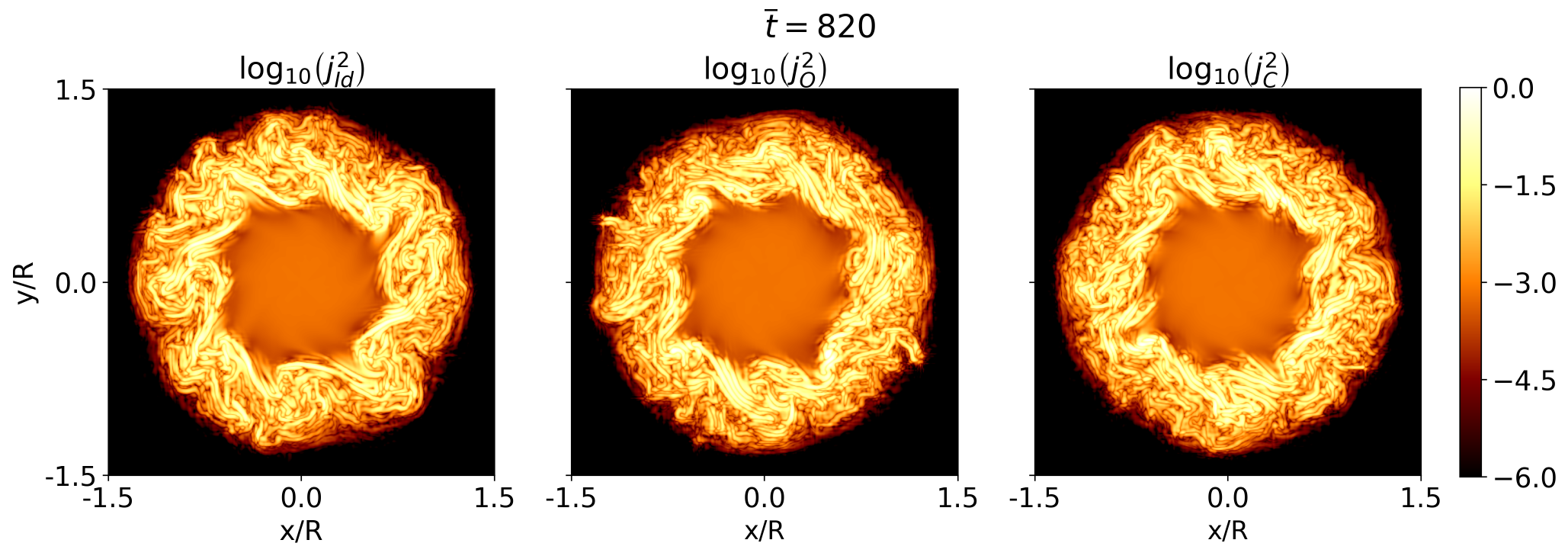}
\caption{Cross-sectional cut at $z=R$ of the current density squared in logarithmic scale at the end of the simulations, $\overline{t}=820 $. The panels are sorted as follows:  ideal simulation (left),  Ohmic simulation (center), and Cowling simulation (right). The complete temporal evolution is available  as an online movie.}
\label{Fig_currentz1triple}
\end{figure*}

The overall, large-scale dynamics of both dissipative simulations are identical to that of the ideal MHD simulation discussed in Sect.~\ref{Sub_idealdyn}. No noticeable differences are seen in the results of density, temperature, and vorticity. We remind the reader that the values of $\eta_{\rm O}$ and $\eta_{\rm A}$ used in the dissipative simulations are the realistic ones in the partially ionized prominence plasma (see Fig.~\ref{Fig_etastempxiinitial}). As a consequence of that, dissipation is not artificially enhanced in our simulations and its influence is necessarily reduced to the smallest scales that develop during the evolution. The current density is the  variable that displays more significant differences between simulations, but even for that variable\textbf{,} the differences can be considered as small. Figure~\ref{Fig_currentz1triple} shows a cross-sectional at $z=R$ of the current density squared in logarithmic scale for the ideal, Ohmic, and Cowling simulations. We find slight differences in the fine scales of the turbulence  that develops at large times in the evolution, but no clear relation can be deduced by comparing the results of the three simulations. It is obvious that the presence of dissipation affects somehow the development of the smallest scales. This fact, together with the intrinsic chaotic nature of turbulence \citep[see, e.g.,][]{Biskmap03} causes the appearance of slightly different turbulent patterns for the current density in the three simulations.

\begin{figure}
\centering
\resizebox{\hsize}{!}{\includegraphics{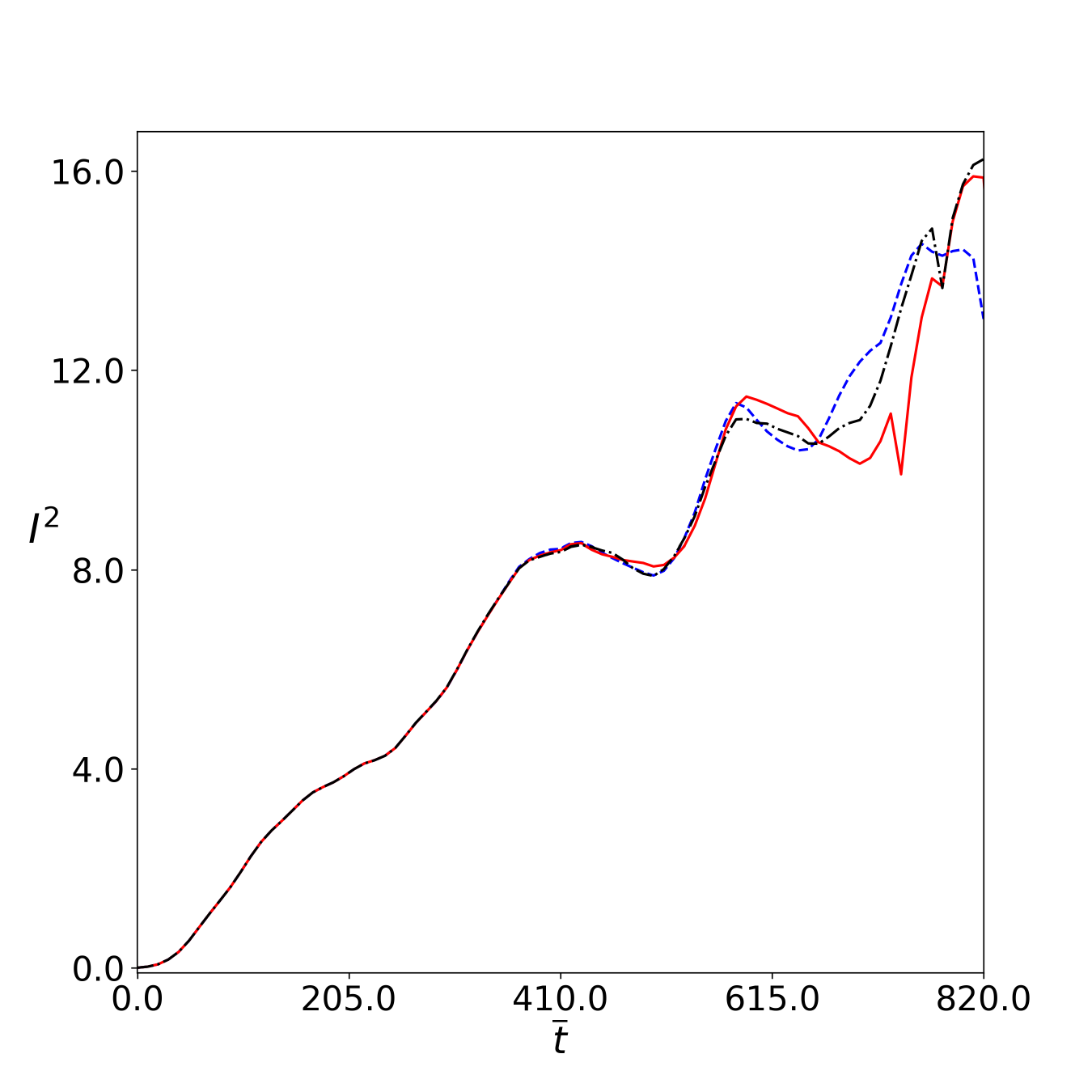}} 
\caption{Same as bottom panel from Fig. \ref{Fig_vortycurry} for the ideal simulation (solid red line), but now including the results from the  Ohmic simulation (blue dashed line) and the Cowling simulation (black dash-dotted line).}
\label{Fig_current_threecases}
\end{figure}

To quantify the differences in the current density between the three simulations, Fig.~\ref{Fig_current_threecases} shows the temporal evolution of the current density squared integrated in the whole computational domain. The curve corresponding to the ideal simulation was already displayed in  the bottom panel of Fig.~\ref{Fig_vortycurry}. The three cases behave similarly until turbulence begins to dominate the dynamics late in the evolution. The three curves separate from that time onwards, although an increasing trend remains in the three cases. This increasing trend points out that the generation of small scales in current density continues even when magnetic diffusion works to dissipate those small scales. Counter-intuitively, the ideal simulation shows lower values of current density than the two dissipative simulations in some time range. In principle, one should expect Ohmic and ambipolar diffusion to slow down the KHI development and inhibit the formation of the smaller KHI vortices compared to the ideal case. In this line of thought, the increase in the current density in the dissipative simulations should be less pronounced than in the ideal simulation. However, the results do not point in that direction. Instead, we find no clear pattern relating the efficiency of physical dissipation with the behavior of the integrated current density. The reason why physical dissipation is  acting inefficiently to damp the perturbations in the current density resides in the realistically small values of the diffusion coefficients and the likely influence of the unavoidable numerical dissipation. Since we are not able to disentangle the role of  numerical dissipation from that of physical dissipation, it is not possible for us to determine the relative importance of numerical dissipation. However, the fact that the three simulations show a different turbulent pattern at the very small scales indicates that  numerical dissipation does not entirely dominates and that physical dissipation is playing a role.

\begin{figure*}
\centering
\includegraphics[width=17cm]{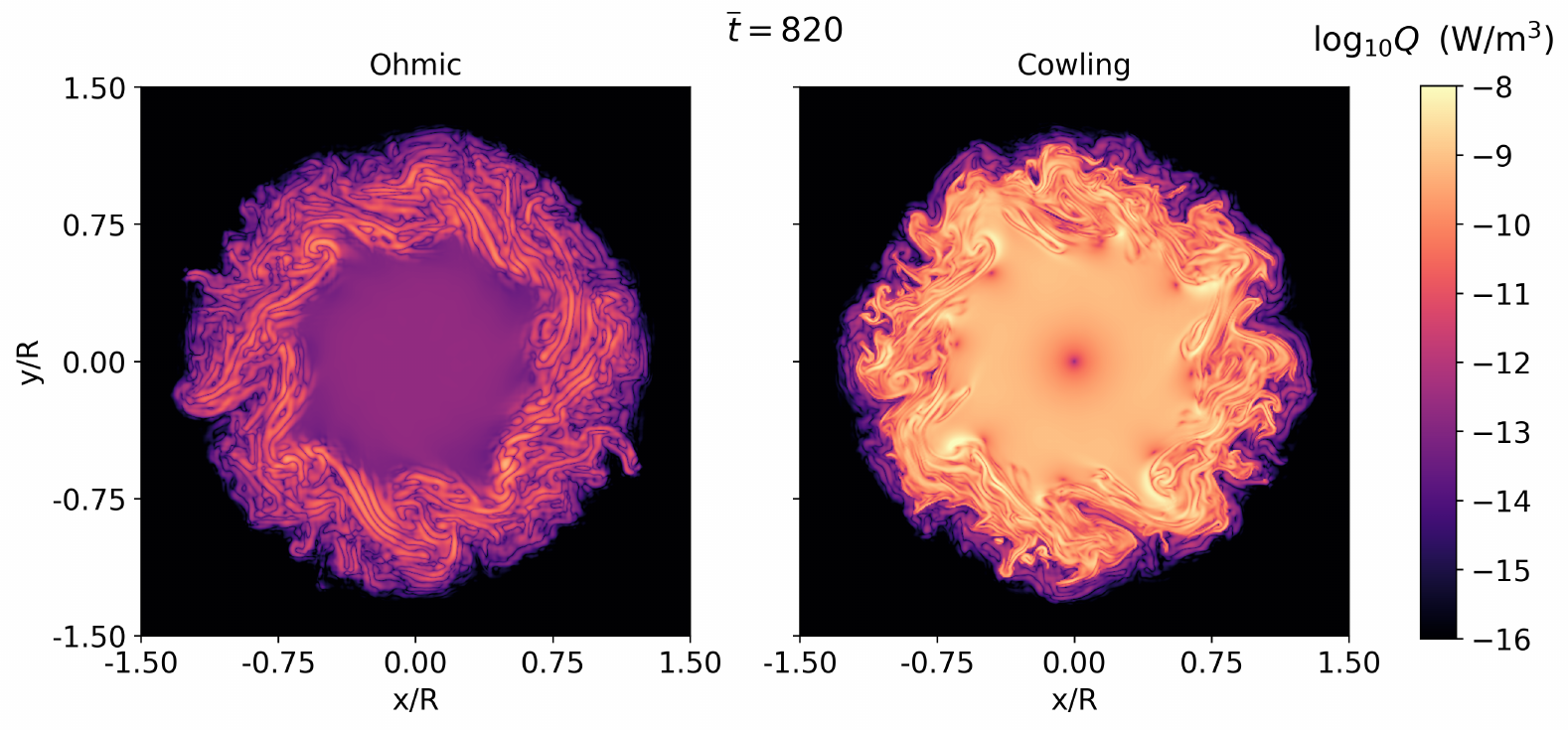}
\caption{Snapshot of the volumetric heating rate, $Q$, in a cross sectional cut at $z=R$ at the end of the simulation time, $\bar{t} = 820$.  A logarithmic scale has been used in both panels for an optimal visualization. \textit{Left panel}:  Ohmic simulation. \textit{Right panel}: Cowling simulation.  The complete temporal evolution is available as an online movie.}
\label{Fig_heating_z1}
\end{figure*}

We go on to examine the heating produced in the dissipative simulations.  Using the approximate resistivity tensor of Eq.~(\ref{etadiagonal}), the volumetric heating rate, $Q$, is:
\begin{equation}
Q= \mu_{0}\left(\hat{\eta} \cdot \mathbf{j}\right) \cdot \mathbf{j} 
\approx \mu_{0}\left(\eta\lvert\mathbf{j}_{\parallel}\rvert^{2}+\eta_{C}\lvert\mathbf{j}_{\perp}\rvert^{2}\right),
\label{heat} 
\end{equation}
where  $ \mathbf{j}_{\parallel} $ and $ \mathbf{j}_{\perp}$ are the components of the current density in the parallel and perpendicular directions to the  background magnetic field, calculated as:
\begin{equation}
\mathbf{j}_{\parallel}=\frac{\mathbf{j}\cdot\mathbf{B}}{\lvert \mathbf{B}\rvert^{2}}\mathbf{B} \approx j_z,
\label{jpara}
\end{equation}
\begin{equation}
\mathbf{j}_{\perp}=\frac{\mathbf{B}\times(\mathbf{j}\times\mathbf{B})}{\lvert \mathbf{B} \rvert^{2}} \approx j_x \hat{x} + j_y \hat{y}.
\label{jperp}
\end{equation}
When ambipolar diffusion is absent, $\eta_{C}=\eta$, and the heating is isotropic. In the presence of ambipolar diffusion, the heating is expected to be more intense and dominated by the dissipation of perpendicular currents. Figure~\ref{Fig_heating_z1} shows the volumetric heating rate in the Ohmic and Cowling simulations at the same cross-sectional cut as that in Fig.~\ref{Fig_current_threecases}.

In the Ohmic simulation, we find that the heating  is negligible in the initial stage of the dynamics governed by phase mixing alone. Once the KHI evolves nonlinearly and turbulence develops, there is a large increase in the values of the heating rate. The heating is essentially produced in the nonuniform transitional layer between the core of the prominence thread and the corona, where turbulence mainly develops. Negligible Ohmic heating is obtained in the cool core of the thread, since turbulence does not completely reach that region of the flux tube. 

The heating obtained in the Cowling simulation is around three orders of magnitude larger than the heating in the Ohmic simulation. Such a result is consistent with the larger efficiency of ambipolar diffusion in prominence threads \citep[see, e.g,,][]{Melis21,Melis23}. Again, the heating is most important in the nonuniform transitional layer, but now a non-negligible heating is obtained in the cool core of the thread. The fact that the cool core is only partially ionized results in the presence of heating although turbulence does not develop in that region. Thus, the heating in the cool core is unrelated to turbulence and is directly caused by the ambipolar dissipation of the Alfv\'{e}n waves \citep{Soler09}.

\begin{figure}
\centering
\resizebox{\hsize}{!}{\includegraphics{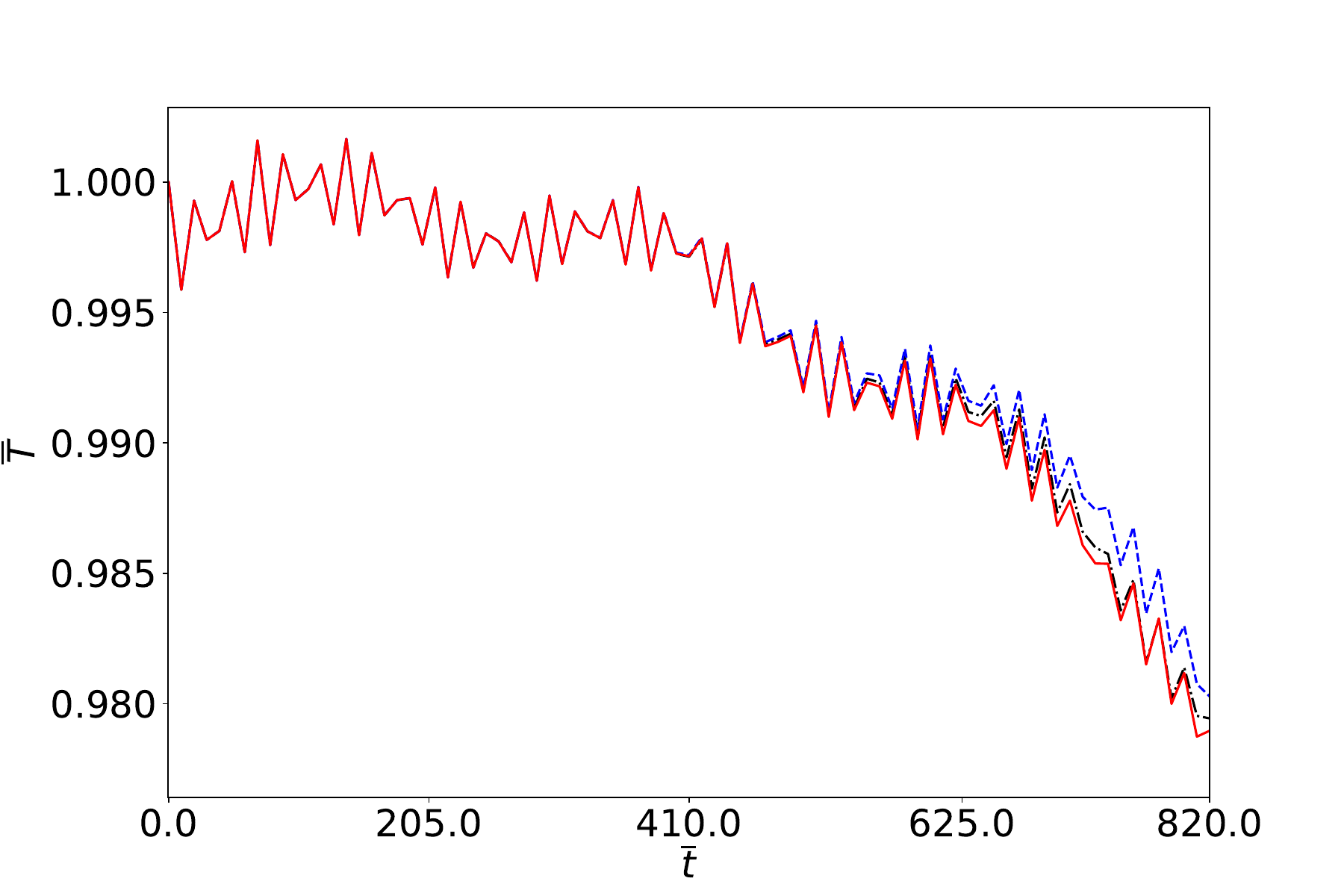}} 
\caption{Temporal evolution of the average temperature in a subdomain where $ x/R \in \left[-1.4,1.4\right]$, $ y/R \in \left[-1.4,1.4\right] $, and $z/R \in \left[-4,4\right] $ for the ideal simulation (solid red line), the Ohmic simulation (blue dashed line) and the Cowling simulation (black dash-dotted line). The temperature is normalized  with respect to the initial average temperature.}
\label{Fig_temperature_threecases}
\end{figure}

The regions where heating is more intense appear to be very spatially localized in the turbulent annulus, which might limit the global effect that this heating could have on the prominence thread. To explore whether the obtained local heating  has a sizeable impact on the thermodynamic state of the thread,  we selected a subdomain of the model that encompasses the partially ionized part of the thread. The considered subdomain corresponds to $ x/R \in \left[-1.4,1.4\right]$, $ y/R \in \left[-1.4,1.4\right] $, and $z/R \in \left[-4,4\right] $. We studied the evolution in time of the average temperature in that region.  The calculation is shown in Fig.~\ref{Fig_temperature_threecases}. For comparison purposes, we also included the result corresponding to the ideal MHD simulation, for which there is no physical heating.  Despite the fact that heating is present in the Ohmic and Cowling simulations,  the average temperature in the three cases shows a similar behavior. During the initial, quasi-linear phase the three curves superimpose. In that phase, we find that the average temperature remains roughly constant, with some slight increases  and decreases probably caused by the adiabatic compression and expansion of the plasma along the flux tube due to the ponderomotive force. When the KHI and the turbulence occur, the  average temperature starts to decrease. The evolution of the average temperature in the dissipative simulations is practically the same as in the ideal simulation in this decreasing phase too.  The average temperature decreases in all cases due to the mixing of the hot coronal plasma and the cold prominence plasma owing to the nonlinear evolution of the KHI and the turbulence, as \citet{Hillier19cool} and \citet{Hillier23} explained. However, we note that unlike in \citet{Hillier23}, in our simulations we do not include radiative losses, which would further cool the plasma until achieving a radiative equilibrium. The decrease of the average temperature that we find in our simulations is exclusively due to the mixing of the internal and external plasmas. At the end of the simulations, the average temperature in the considered subdomain has decreased about 2\% with respect to the initial value, so that this apparent cooling of the thread due to the plasma mixing is indeed quite modest, but still much more important that the Ohmic and ambipolar heating.   We found a similar evolution for the average temperature using larger subdomains in $z$- $x$- and $y$-directions, as long as the mixing layer is included (not shown here).

We conclude that the plasma heating in the dissipative simulations does not play a relevant role in the thermodynamic state of the thread. The heating is very localized in a turbulent annulus around the cool core of the thread and its contribution is completely overwhelmed by the effective cooling caused by the plasma mixing. The results here are in apparent contradiction to those of  \citet{Melis21}, who found that heating by Alfv\'{e}n waves in prominence threads could be  important enough to balance energy losses due to radiative cooling. However, there are important differences between the work of \citet{Melis21} and the present work, so that their results cannot directly be compared with our findings. \citet{Melis21} studied propagating waves in a frequency range far higher than in our work (they considered frequencies as high as 1 Hz), while here we excited standing  waves that have much lower frequencies (between 1.22 mHz and 6.54 mHz). The efficiency of Alfv\'en wave dissipation increases with the frequency, so that the propagating waves studied by  \citet{Melis21} are more efficiently damped than the standing modes studied here. In addition, \citet{Melis21} used a 1D thread model and neither the KHI nor the turbulence are able to develop in their configuration.

\section{Forward modeling}
\label{sec:forward}

Here we explore the possible observational signatures that the nonlinear evolution of torsional Alfv\'{e}n oscillations in prominence threads may leave. To this end, and inspired by the work of \citet{Martinez22} in the case of kink oscillations, we computed synthetic H$\alpha$ observations, 

We used the method of fast synthesis of the H$\alpha$ line profile given in  \citet{Heinzel15}. The details of the method can be found in \citet{Heinzel15} and we only give a summary of its implementation. The formal solution to the radiative transfer equation in the line-of-sight (LOS) direction is:
\begin{equation}
I_{\nu}=S\left[1-\exp{\left(-\tau_{\nu}\right)}\right],
\label{spec_i}
\end{equation}
where $S$ is the  source function assumed uniform and $\tau_{\nu}$ is the optical thickness in the LOS, defined as:
\begin{equation}
\tau_{\nu}=\int_{LOS} \kappa(\nu, l) \mathrm{d}l,
\label{tau}
\end{equation}
where $l$ is the LOS coordinate and $\kappa(\nu, l) $ is the absorption coefficient given by
\begin{equation}
    \kappa(\nu, l) = \frac{\pi e^{2}}{m_{e}c}f_{23}n_{2} (l)\;\phi\left(\nu,l\right). \label{eq:kapppa}
\end{equation}
In Eq. (\ref{eq:kapppa}), $f_{23} $ is the H$ \alpha$ line oscillator strength \citep[see,][for the tabulated value]{Goldwire68}, $c$ is the speed of light, and $m_{e}$ is the electron mass. Moreover, $ n_{2} $ is the population of hydrogen atoms in the second quantum level and $\phi\left(\nu,l\right) $ is the normalized absorption profile, which is approximated by a Gaussian profile  \citep[see][]{Heinzel15}. From here on, we assume that the LOS direction is the $y-$direction, so the $x-$ and $z-$directions form the plane of sky (POS). Bearing in mind the local Doppler shifts owing to the LOS velocity in the $y-$direction, we write,
\begin{eqnarray}
\phi\left(\nu,y \right)=\frac{1}{\sqrt{\pi}\Delta \nu_{D}(y)}\exp\left\lbrace- \frac{\left[\nu-\nu_{0}\left(1+v_{y}(y) /c\right)^{-1}\right]^{2}}{\Delta \nu_{D}^{2}(y)}\right\rbrace,
\label{normabsorp}
\end{eqnarray}
where $\nu_0$ is the H$\alpha$ rest frequency, $v_{y} (y)$ is the $y-$component of velocity, and $\Delta \nu_{D} (y)$ is the thermal and microturbulent broadening given in \citet{Heinzel15}. The microturbulent broadening depends on the microturbulent velocity, whose value is set to 5~km~s$^{-1}$ as in \citet{Heinzel15}. In the simulations, we occasionally found maximum LOS velocities of $ \pm $ 10 km/s, which is the limit of accuracy of the method of \citet{Heinzel15}. However, the typical LOS velocities are normally around $ \pm $ 6 km/s, which are compatible with observationally reported values \citep[see, e.g.,][]{Schmieder10,Gunar12}.

The hydrogen second-level population, $ n_{2}$, can be calculated from the electron density. Following \citet{Heinzel15}, both physical quantities are related as,
\begin{equation}
n_{2} (y)=\frac{n^{2}_{e}(y)}{f(T(y),p(y))},
\label{popsec}
\end{equation}
where $f(T(y),p(y))$ is a function that depends on temperature and gas pressure. Physically, the function $f$ is associated with the rate of the photoionization and the radiative recombination from and to the hydrogen second-level population \citep{Heinzel94}. We calculated the values of the function $f$ using bilinear interpolation from the values in Table 1 in \citet{Heinzel15}, assuming a constant altitude of 10 Mm. We are aware that the source function can vary with height \citep[see, e.g.,][]{Heinzel94}. However, for consistency with our choice, we neglected the variation of the source function with height.

\begin{figure*}[htbp!]
\centering
\includegraphics[width=17cm]{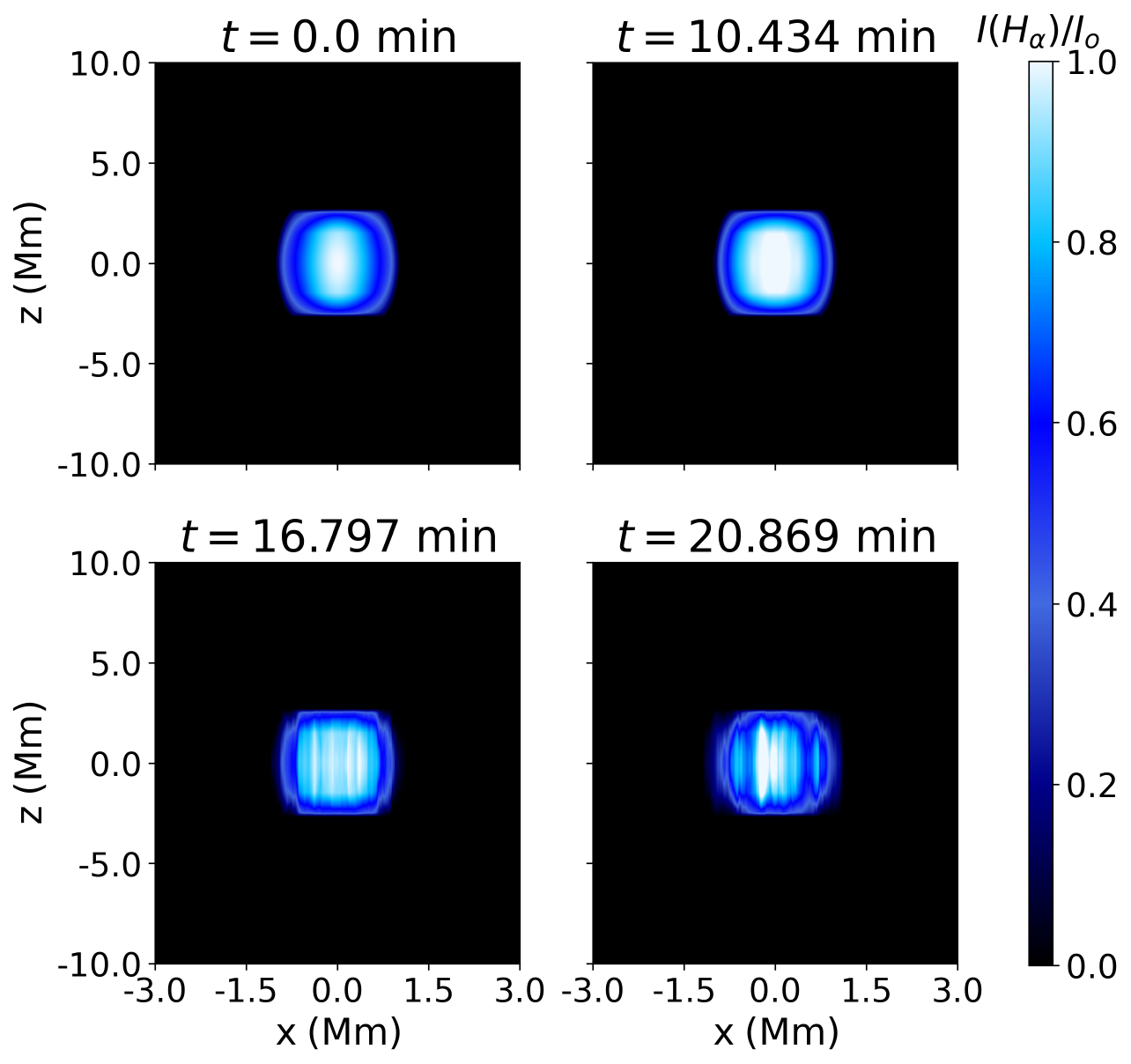}
\caption{Normalized specific intensity at the H$\alpha$ rest frequency in four different times. The normalization is done with respect to the maximum value at the beginning of the simulation. We note that the longitudinal, $z$, and transverse, $x$, directions to the thread are not to scale. A complete temporal evolution is available as an online movie.}
\label{Fig_zoomed}
\end{figure*}

\begin{figure*}
\resizebox{0.45\hsize}{!}{\includegraphics{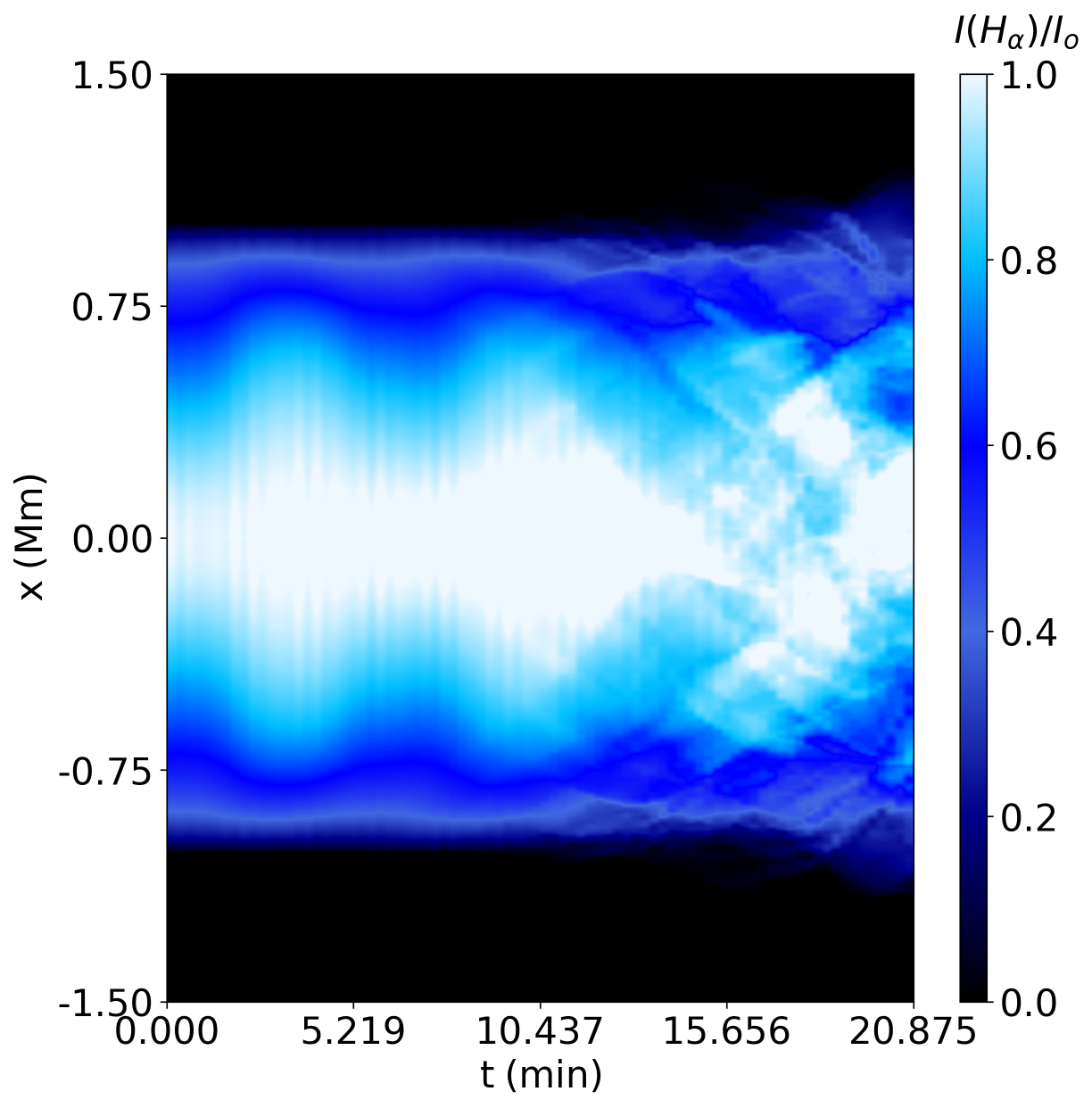}} 
\resizebox{0.45\hsize}{!}{\includegraphics{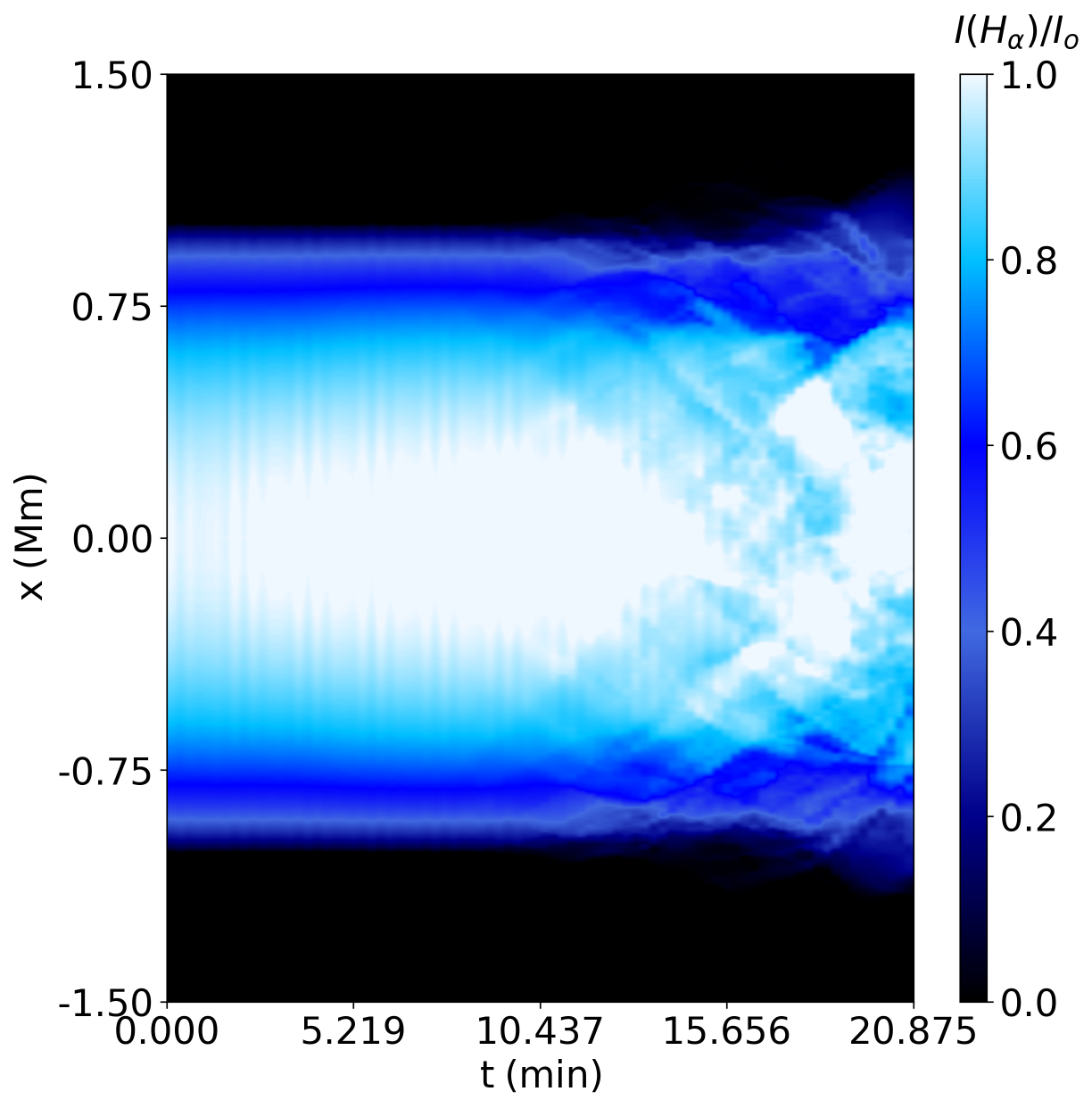}} 
\centering
\caption{Time-distance map of the normalized specific intensity across the thread at the rest frequency of H$\alpha $ (left). The normalization is done with respect to the maximum value at the beginning of the simulation. Right panel is the same as the left, but neglecting the LOS velocities.}
\label{Fig_timedistance}
\end{figure*}

Since the ideal and dissipative simulations are nearly identical in the description of the overall dynamics of the thread, we only use here the results from the ideal simulation to calculate the synthetic observation. The specific intensity at the H$\alpha$ rest frequency in the POS is displayed in Fig.~\ref{Fig_zoomed} and its accompanying animation. As  the fully ionized plasma does not emit in H$\alpha$,  we used a reduced numerical domain in $z$ where $z/R \in [-10,10]$. The  emission occurs at the core of the partially ionized prominence thread, which is the only part of the model that is visible in the synthetic images. In the temporal evolution, first  one can see a periodic pulsation in the intensity that emerges from the thread. This initial phase is then followed by the appearance of fine strands that are longitudinal to the thread axis and have a brighter intensity than the rest of the thread.  The reason for the existence of these two distinct phases in the synthetic observation is analyzed below.

The initial periodic variations in the H$\alpha$ intensity are caused by the integration of the LOS flows and, to a lesser extent, the ponderomotive flows along the thread.  To confirm this hypothesis, first, we created a time-distance map of the H$\alpha$ intensity at  $z=0$, that is, across the thread. This is displayed in Fig.~\ref{Fig_timedistance} (left panel). The initial intensity pulsations are clearly visible. The period of such pulsations is $\approx 6.6$~min and roughly two periods are seen in the time-distance map. Then, we repeated the time-distance map but neglecting the LOS flows. This is displayed in Fig.~\ref{Fig_timedistance} (right panel). The comparison of the two maps reveals that the LOS flows or, more precisely, the integration of those flows along the LOS, have the net effect of producing the intensity pulsations. In the initial stages of the evolution, the LOS flows are caused by the phase mixing of the Alfv\'en modes. Therefore, the intensity pulsations appear to be an observational manifestation of phase mixing.  To the best of our knowledge, this has not been discussed before in the literature.

Simultaneously to the pulsations, one can see a gradual brightening at the core of the prominence thread. This brightening is present in both panels of Fig.~\ref{Fig_timedistance}, although it is visually more evident in the right panel where the pulsations are absent. This gradual brightening is caused by the ponderomotive force, which tends to accumulate plasma around the thread center so increasing its density. Using Eq.~(14) from  \citet{diazsoler21aanda}, we obtained the period of the associated slow MHD wave due to the ponderomotive force to be  $ \approx 24.5 $~min. To compute this period, we considered the longitudinally averaged density along the axis of the flux tube. The ponderomotive force acts so slowly that the KHI develops before a complete period of the slow MHD wave. In the time-distance map, the KHI  and the onset of turbulence are seen in the form of an irregular pattern of bright patches ultimately caused by the development of the KHI vortices and the mixing of plasma. We note that the omission of the LOS flows does not alter much the structure of the bright patches, which confirms that their origin resides in the turbulent distribution of the density.

By the time that the KHI develops after $\sim 12$~min, Fig.~\ref{Fig_zoomed} shows the formation of fine strands at the core of the prominence thread. These strands are the observational signature of the  KHI vortices, which develop perpendicularly to the tube axis. This is not a new result. The formation of strands owing  to the KHI has been found in numerical simulations of magnetic tubes oscillating in the kink mode \citep[e.g.,][]{Antolin14,Antolin15,Antolin16,Antolin17,Guo19,Shi21,Martinez22}. Most of the previous works performed forward modeling of coronal loops, which involves spectral lines associated to much hotter temperatures than those of prominence threads.

\citet{Martinez22} studied the H$\alpha$ emission of prominence threads oscillating in the kink mode and our results can be compared to theirs. Unlike in \citet{Martinez22}, we do not see a global lateral oscillation of the flux tube, which is characteristic of kink modes. Another difference is that  the initial periodic pulsations inside the prominence thread due to the phase-mixing flows along the LOS  are not reported in \citet{Martinez22}. Regarding the formation of the strands, \citet{Martinez22} found that the width of the  strands in their kink mode simulations is between 10 km and 125 km. We found that the width of the strands in our torsional mode simulations range between 115 km and 168 km, approximately. Considering such size of the strands, they might be observable with instruments such as CRisp Imaging SpectroPolarimeter (CRISP; \citeauthor{Scharmer08CRISP} \citeyear{Scharmer08CRISP}) at the Swedish 1 m Solar Telescope (SST; \citeauthor{Scharmer03SST} \citeyear{Scharmer03SST}), or the Visible Image Spectrometer (VIS, \citeauthor{Cao10NST} \citeyear{Cao10NST}) at the Goode Solar Telescope (GST, \citeauthor{Goodsolartelescope} \citeyear{Goodsolartelescope}). Both instruments has an spatial resolution approximately equal to 0.1" ($ \sim 70 $ km). 

However, we should note that the radius of our simulated thread, 1~Mm, is larger than the typically reported thread radii, $\sim 200-300$~km  \citep[see, e.g.,][]{Lin05,Lin07}. An average value of observed thread radii is 228~km \citep[see, e.g.,][]{Arregui18LRSP}, which is a factor of 4.38 smaller than the radius of the simulated thread. After scaling the width of the strands by the same factor, we obtain that the strands should have an actual width between 26 km and 38 km, approximately. Such fine strands cannot be seen with current instruments. Conversely, the next generation of solar telescopes might be able to observe these  fine strands. The Visible Broadband Imager (VBI; \citeauthor{Wogervbi21} \citeyear{Wogervbi21}) and the Visible Tunable  Filter (VTF; \citeauthor{Schmidtvtf14} \citeyear{Schmidtvtf14}), both installed at \textit{Daniel K. Inouye} Solar Telescope (DKIST; \citeauthor{dkist20} \citeyear{dkist20}), have spatial resolutions of 20 km and $ \sim 25 $ km, respectively. The Tunable Imaging Spectropolarimeters (TIS) installed at the European Solar Telescope (EST; \citeauthor{Quintero22} \citeyear{Quintero22}) have an spatial resolution of $ \sim 29 $ km at the rest wavelength of H$\alpha$ in air.
We repeated the synthetic modeling considering different LOS directions, but all cases showed similar results.

\section{Concluding remarks}
\label{sec:conclusions}

In this numerical work, we study the nonlinear evolution of standing torsional Alfv\'{e}n waves in a low-$\beta$ prominence-thread model embedded in a constant and axial magnetic field. The model consists of a flux tube that has a core, a transversely nonuniform transition region, and a external medium. The external medium has a uniform temperature and density while the longitudinally nonuniform core of the prominence thread is denser and cooler than the external medium. Both regions are continuously connected everywhere through a nonuniform transition. Moreover, the magnetic field, where the prominence thread is embedded, is line-tied at the photosphere. For simplicity, no chromospheric layer is included. 

We excited the longitudinally fundamental mode of standing torsional Alfv\'{e}n waves and performed three simulations using the PLUTO code: an ideal MHD simulation and two nonideal simulations including Ohmic diffusion alone and Ohmic together with ambipolar diffusion. Other nonideal effects, such as the Hall effect and the Biermann battery effect, were ignored as they are of much less relevance in prominences \citep{Khomenko14b,Ballester18,Melis21}. The simulated dynamics undergoes three differentiated stages. The first stage is quasi-linear and dominated by the phase mixing of the Alfv\'en modes \citep[see, e.g.,][]{HeyvaertPriest83}. The second stage begins when phase-mixing azimuthal flows trigger the KHI. Since the KHI excites higher azimuthal modes, the KHI onset time has been  obtained through an analysis of Fourier modes in the azimuthal direction using the azimuthal and radial components of the velocity at a specific radial position.  The onset time obtained from the Fourier analysis agrees with the time at which a change of trend in the integrated values of vorticity happens. Once the KHI is triggered,  large eddies initially appear, which evolve nonlinearly breaking into smaller and smaller eddies. This dynamics eventually leads to turbulence, so the  dynamics of the thread is similar to that obtained in simulations of torsional oscillations of coronal loops \citep[see][]{diazsoler21aanda}. Nonetheless, the limited spatial resolution of our ideal MHD simulation imposed a third phase, namely the saturated phase. In this stage, the integrated  vorticity remains roughly constant when it should increase indefinitely. This situation, however, does not impede that the generation of turbulence is captured by our simulations.

Overall, the large-scale dynamics in the two dissipative simulations is nearly identical to that of the ideal simulation. Some differences appear only at the very small scales, specially in the current density. This   rather subtle effect of dissipation, indeed, causes some impact in the evolution of turbulence, but we find it difficult to disentangle the effect of the physical dissipation from that of the inherent numerical dissipation. The plasma heating associated with the dissipation of currents is weak and very localized in an annulus region at the thread boundary. As a result of that, no global heating of the prominence thread is produced. Instead, the mixing of the cool prominence plasma with the hot coronal plasma leads to a moderate cooling of the thread towards an intermediate temperature \citep[see][]{Hillier23}. Thus, Ohmic and ambipolar dissipation of turbulence induced by torsional waves does not appear to be a mechanism capable of heating prominence threads.

We wondered what observational signatures the KHI and the turbulence  might leave. To this end, we followed the method of \citet{Heinzel15} to compute synthetic H$\alpha$ observations.  Before the onset of the KHI, we found a periodic brightening of the H$\alpha$ intensity at the core of the prominence thread. This periodic brightening is caused by the integration of the flows along the LOS and, to a lesser extent, the longitudinal flows driven by the ponderomotive force. Later, we saw the formation of fine bright strands parallel to the thread axis owing to the KHI vortices and the turbulence. Forward modeling of coronal loop kink oscillations \citep[see, e.g.,][]{Antolin14,Antolin16,Antolin17,Guo19,Shi21} and prominence thread kink oscillations \citep{Martinez22} reported similar strand formation. We showed that standing torsional oscillations are also able to drive such fine strand formation. When scaled according to the considered radius of the thread in the simulations, the effective width of the strands ranges between 26 km and 38 km, approximately. The next generation of solar telescopes DKIST \citep{dkist20} and EST \citep{Quintero22} might be able to observe them. We note that while we modeled a prominence thread as a straight magnetic tube, in reality the prominence material is deposited in dips of the magnetic field due to the effect of gravity. Thus, their inclusion in future works would be more realistic. The study of the interaction between neighboring threads in a prominence would also be an interesting extension of this work.


\begin{acknowledgement}
This publication is part of the R+D+i project PID2020-112791GB-I00, financed by MCIN/AEI/10.13039/501100011033. S.D.S. acknowledges the financial support from MCIN/AEI/10.13039/501100011033 and European Social Funds for the predoctoral FPI fellowship PRE2018-084223. This research was supported by the International Space Science Institute (ISSI) in Bern, through ISSI International Team project \#457 (The Role of Partial Ionization in the Formation, Dynamics and Stability of Solar Prominences). We acknowledge the UIB for the use of the Foner cluster. S.D.S thanks L. Melis, M. Kriginsky, and F. Bail\'{e}n Mart\'{i}nez for their help. We also thank the anonymous referee for the constructive comments that improved the quality of the paper. During the analysis of data, we have used VisIT \citep{VisIt}, Mathematica \citep{Mathematica}, and Python 3.6. The Python packages that we have used are Matplotlib \citep{Hunter07}, Scipy \citep{Virtanen20} and Numpy \citep{Harris20}. We are thankful to B. Vaidya and his contributors for the tool pyPLUTO.
\end{acknowledgement}

\bibliographystyle{aa} 
\bibliography{thebiblio} 

\begin{appendix}

\section{Justification for neglecting off-diagonal terms in the resistivity tensor}
\label{app:approx}

\begin{figure}[htbp!]
\centering
\resizebox{\hsize}{!}{\includegraphics{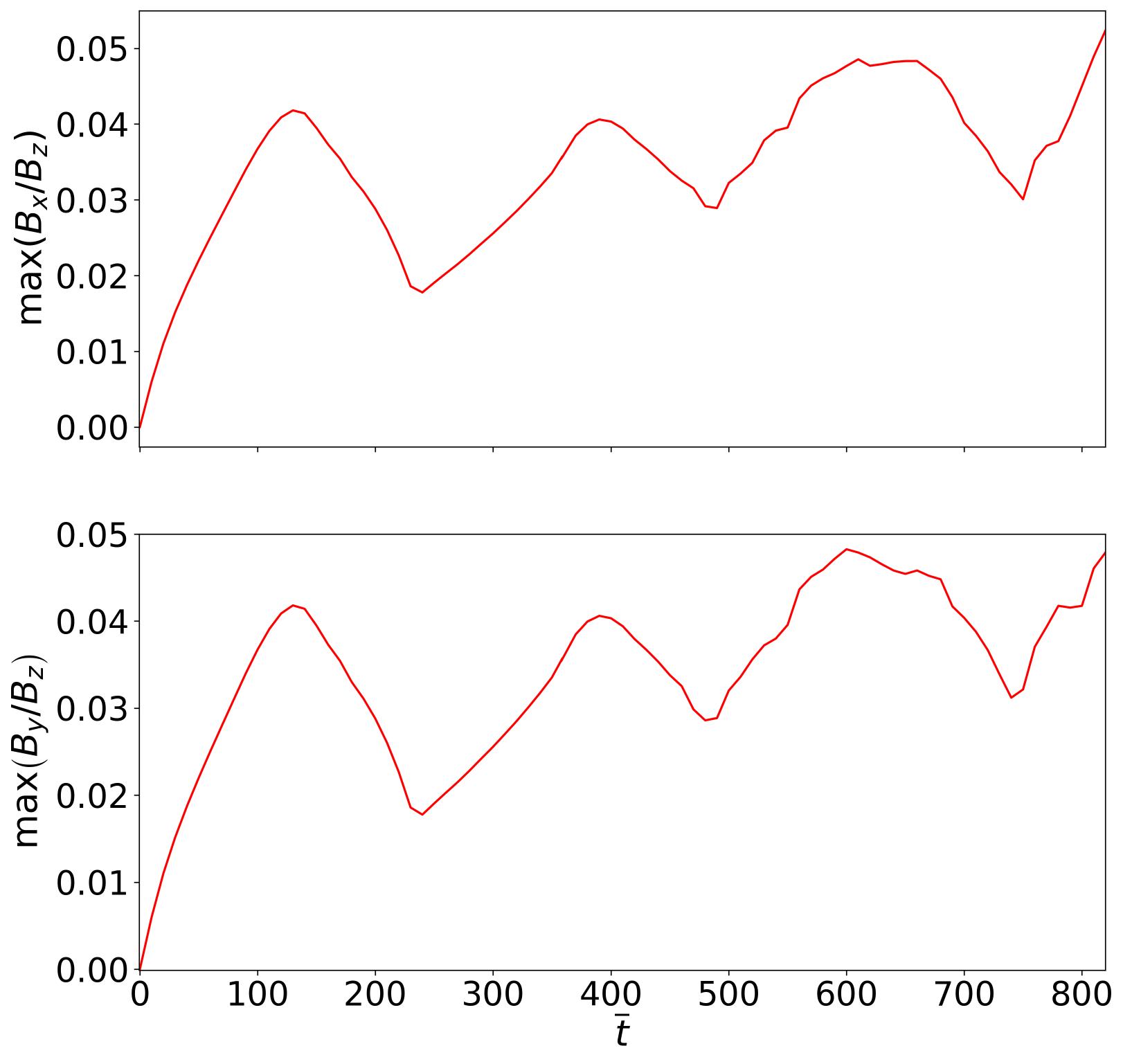}} 
\caption{Temporal evolution of the maximum value of  $B_x/B_z$ (top). The same, but for $B_y/B_z$ (bottom). The results are obtained from the ideal MHD simulation. }
\label{Fig_maxbxbz_maxbybz}
\end{figure}

As mentioned in Sect.~\ref{nonideal}, the resistivity tensor is defined in PLUTO as a diagonal tensor (see Eq.~(\ref{eq:etatensor})). However, the ambipolar diffusion introduces off-diagonal elements, as shown in Eq.~(\ref{etahat}). Nonetheless, the properties of our problem allows us to implement the ambipolar diffusion in an approximate manner. Particularly, the background magnetic field, that is aligned with the main axis of the flux tube along the $z$-direction, is much stronger than the components across the background magnetic field, namely $B_x$ and $B_y$. This observation allows us to neglect the off-diagonal elements associated with ambipolar diffusion. For the ideal MHD simulation,  we computed the maximum values of $B_x/B_z$ and $B_y/B_z$ and plotted them against time. These results are displayed in Fig.~\ref{Fig_maxbxbz_maxbybz}. As one can see, the maximum values of the  magnetic field components  perpendicular to the background  field are  less than 5\% the value of the background  field strength. In addition, we checked that these maximum values typically occur near the footpoints of the magnetic tube, where the plasma is fully ionized and  ambipolar diffusion is absent. Therefore, the approximation explained in Sect.~\ref{nonideal} to implement the ambipolar diffusion in the particular case under study is justified.

\section{Test implementation of  Ohmic and ambipolar diffusion in PLUTO}
\label{AppendixB}

\subsection{For 1.5 dimensions}
\label{app1}

Following \citet{Balsara96}, \citet{Soler09}, or  \citet{Ballester18}, the dispersion relation for parallel propagating linear Alfv\'{e}n waves in the presence of the Cowling (Ohmic + ambipolar) diffusion is: 
\begin{equation}
\omega^{2}+i k^{2} \eta_{\rm C}\omega-k^{2}v^{2}_{\rm A}=0,
\label{dispe}
\end{equation}
where $\omega$ is the frequency, $ k$ is the parallel wavenumber, and $v_{\rm A}$ is the Alfv\'{e}n speed. The solution of Equation (\ref{dispe}) is: 
\begin{equation}
    \omega = \pm k v_{\rm A} \sqrt{1-\frac{k^2 \eta_{\rm C}^2}{4 v^{2}_{\rm A}}} - i \frac{k^2 \eta_{\rm C}}{2}. 
\label{eq:theorfreq}
\end{equation}
The complex part of the frequency is  related to the damping owing to Cowling's diffusion.

In this test, we considered a 1D domain in the $x-$direction with $x\in[0,L]$ and with the same physical conditions as in the core of the prominence thread. The length of the domain is set to $L = 10 \;\mathrm{km}$ to consider a short wavelength and so to enhance the role of diffusion.  The magnetic field is uniform and aligned with the $x-$direction as well. The fundamental standing  Alfv\'{e}n wave is excited  by considering an initial condition for the $y$-component of the velocity as
\begin{eqnarray}
    v_y = v_0 \sin\left( \frac{\pi x}{L}\right),
\end{eqnarray}
with $v_0 = 0.01 v_{\rm A}$ to be in the linear regime. Regarding the boundary conditions, we set all the variables to satisfy outflow conditions, that is, zero gradient, except for the three components of velocity, which are fixed to zero. We used a numerical grid of 1000 points.

\begin{figure}
\centering
\resizebox{\hsize}{!}{\includegraphics{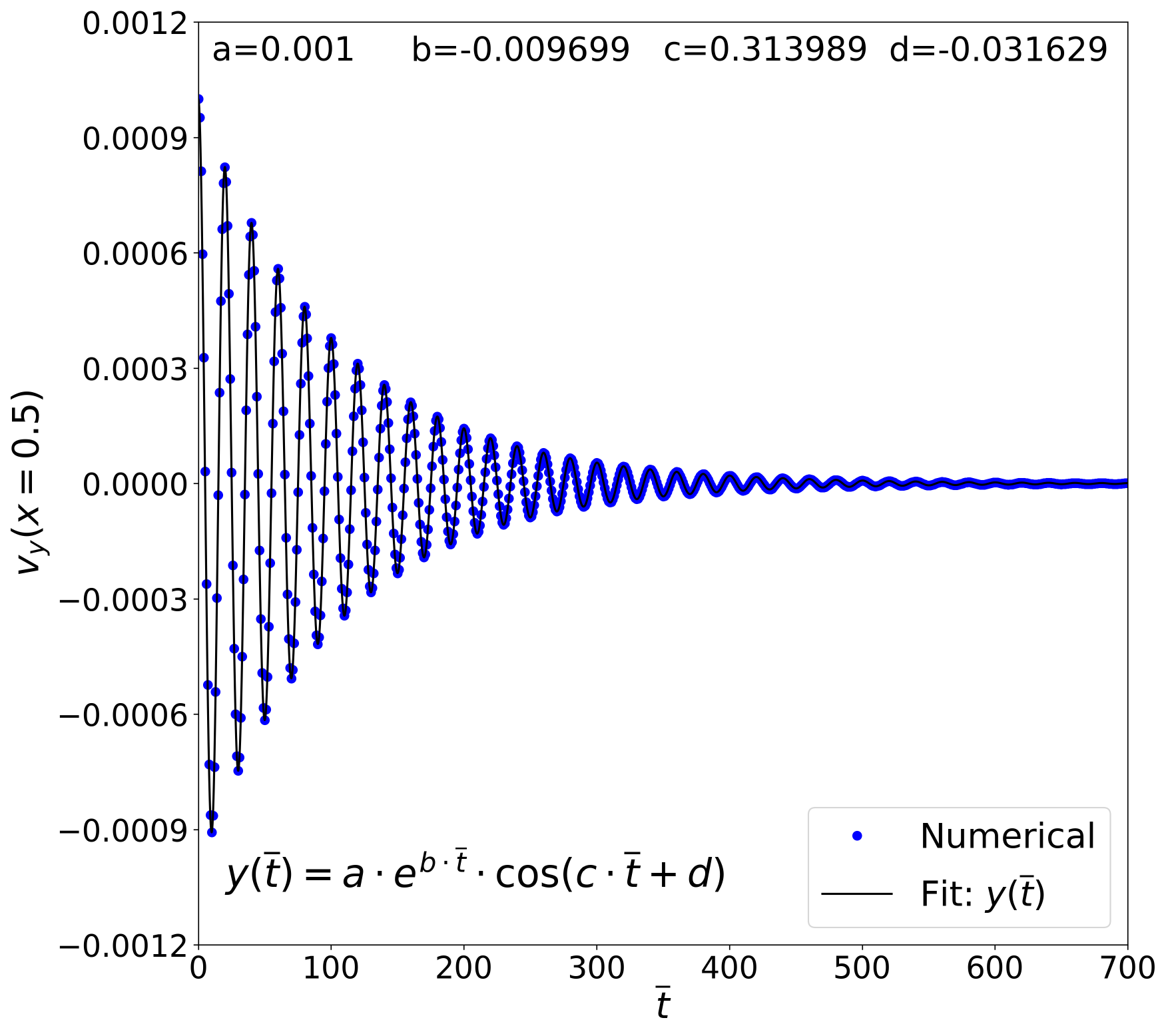}} 
\caption{Temporal evolution of the $y$ component of velocity at the anti-node. The blue dots corresponds to the results from the 1.5D simulation while the black solid line corresponds to the fitting of the data by a damped sine function, as indicated in the plot. Code units are used.}
\label{fig_cowling15}
\end{figure}

We studied the damping of the Alfv\'{e}n wave by fitting the maximum value of the $y-$component of velocity with an exponentially damped sine function. The result is shown in Fig. \ref{fig_cowling15}. We normalized all quantities using the dimensional values of $L$ and $v_{\rm A}$. Then, in dimensionless values $k = \pi$ and $\eta_{\rm C} \approx 1.96\times 10^{-3}$. Using Eq.~(\ref{eq:theorfreq}), the theoretical frequency is $\omega \approx 0.314010-i \;0.009672 $. In turn, the frequency obtained from the simulations is $\omega \approx 0.313989 -i \;0.009699 $. Therefore, we find that the simulations correctly recover the theoretical frequency.

\subsection{For 2.5 dimensions}
\label{app2}

We solved the diffusion problem of the magnetic field in 2D under the physical conditions of the core of the prominence thread. The velocity components are all set to zero, so the problem is solved under static conditions. This test is inspired by that included in the PLUTO documentation to verify the implementation of the resistivity module\footnote{\url{http://plutocode.ph.unito.it/Doxygen/Test_Problems/_m_h_d_2_resistive___m_h_d_2_field___diffusion_2init_8c.html\#details}}. 

We solved the problem in a uniform grid of 512x512 points where $x,y \in [-L,L]$, with $ L=10 \;\mathrm{km}$. We considered a uniform background magnetic field in the $z$-direction with strength $B_{0}$. Outflow  boundary conditions are used.

Again, we normalized all quantities using the dimensional values of $L$ and $v_{\rm A}$. We considered the approximate form of the resistivity tensor given in  Eq.~(\ref{etadiagonal}).  The simulation is initiated at $\bar{t}=1$ with the following prescription for the components of the magnetic field:
\begin{eqnarray}
B_{x}(\bar{t}=1) &=& \varepsilon\exp{\left(-\frac{y^{2}}{4\eta_{\rm O}}\right)}, \label{bx2d} \\ 
B_{y}(\bar{t}=1) &=& \varepsilon\exp{\left(-\frac{x^{2}}{4\eta_{\rm O}}\right)}, \label{by2d} \\ 
B_{z}(\bar{t}=1) &=& B_{0}+ \varepsilon\exp{\left[-\frac{x^{2}+y^{2}}{4\eta_{C}}\right]}, \label{bz2d}
\label{bequations2dinitial}
\end{eqnarray}
with $\varepsilon = 0.001 B_0$, so that $B_x, B_y \ll B_z$ to be consistent with our approximate implementation of the ambipolar diffusion. The temporal evolution of the magnetic field for $\bar{t}>1$ can analytically be obtained as 
\begin{eqnarray}
B_{x}(\bar{t}) &=& \frac{\varepsilon}{\sqrt{\bar{t}}}\exp{\left(-\frac{y^{2}}{4\eta_{\rm O} \bar{t}}\right)}, \label{bx2dtno0} \\ 
B_{y}(\bar{t})  &=& \frac{\varepsilon}{\sqrt{\bar{t}}}\exp{\left(-\frac{x^{2}}{4\eta_{\rm O} \bar{t}}\right)}, \label{by2dtno0} \\ 
B_{z}(\bar{t})  &=& B_{0}+ \frac{\varepsilon}{\bar{t}}\exp{\left[-\frac{x^{2}+y^{2}}{4\eta_{C}\bar{t}}\right]}, \label{bz2dtno0}
\label{bequations2d}
\end{eqnarray}

\begin{figure}
\centering
\resizebox{\hsize}{!}{\includegraphics{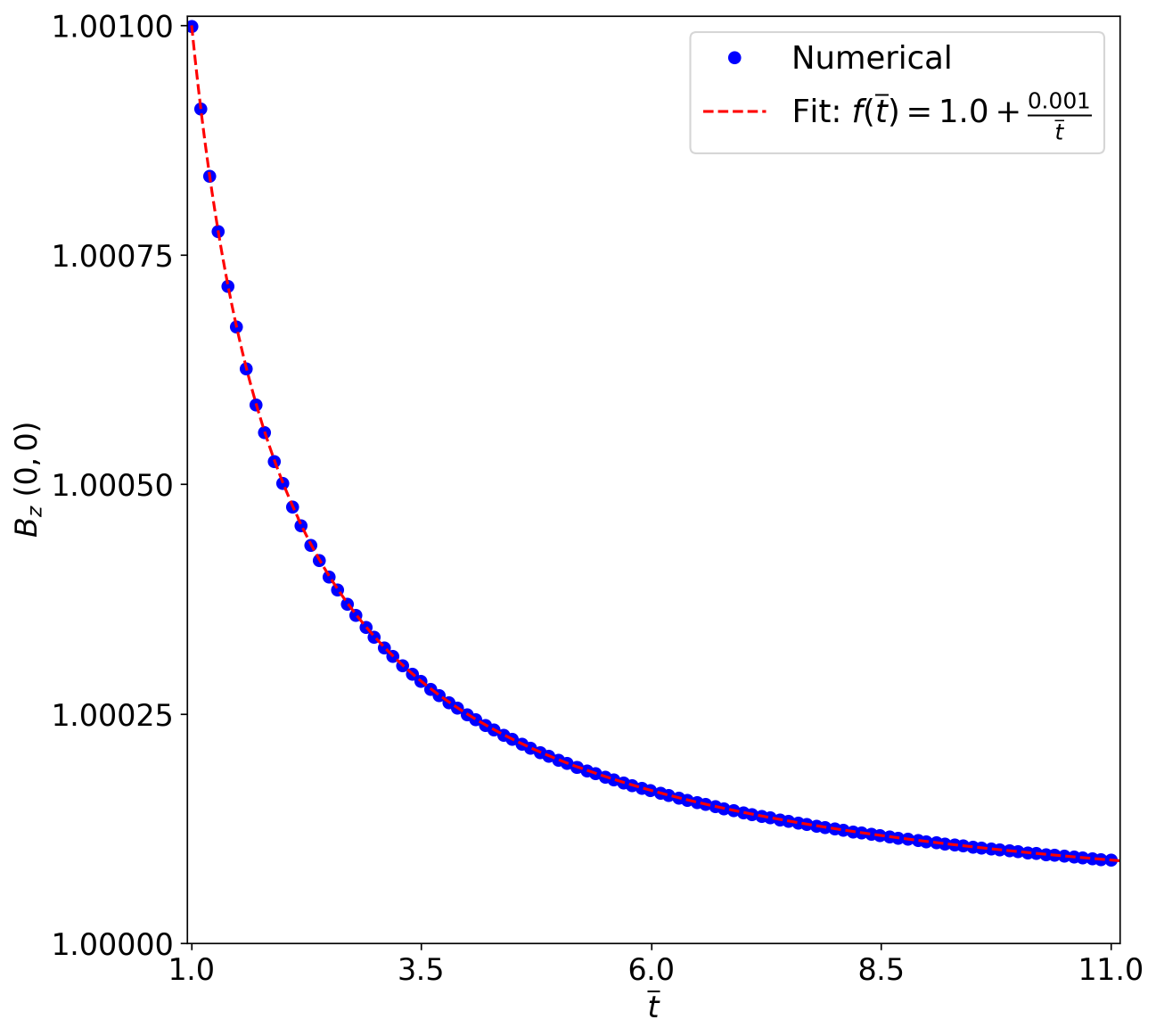}} 
\caption{Temporal evolution of the $z-$component of the magnetic field at the center of the numerical domain. The blue dots corresponds to the results from the 2.5D simulation while the red dashed line corresponds to the fitting of the data by the function: $ f(\overline{t})=1.0+1.0/\overline{t} $. Code units are used.}
\label{Fig_cowling_2D}
\end{figure}

As a verification of the test, we plotted in Fig.~\ref{Fig_cowling_2D} the evolution of the $z-$component of the magnetic field at the center of the numerical domain, $x=y=0$, obtained from the numerical simulation. According to Equation~(\ref{bz2dtno0}), $B_z$ should decrease as $1/\bar{t}$ in that point. The numerical results agree perfectly with the expected dependence. We have verified that the other components of the magnetic field also follow the analytical result (not shown here), meaning that the diffusion problem of the magnetic field is correctly solved in 2D in a situation with $B_x, B_y \ll B_z$.

\subsection{For three dimensions}
\label{app3}

As in the Appendix~\ref{app2}, we solved the diffusion problem of the magnetic field under the physical conditions of the core of the prominence thread, but now in a 3D domain. We used 128x128x128 points in a uniform grid where $x,y,z \in [-L,L]$,  with $ L=10 \;\mathrm{km}$ as before. The situation is equivalent to that solved in the Appendix~\ref{app2} but adding the third dimension.

The simulation is initiated at $\bar{t}=1$ with the following prescription for the components of the magnetic field:
\begin{eqnarray}
B_{x}(\bar{t}=1) &=& \varepsilon\exp{\left(-\frac{y^{2}}{4\eta_{\rm O}}\right)}\exp{\left(-\frac{z^{2}}{4\eta_{\rm C}}\right)}, \label{bx3d} \\ 
B_{y}(\bar{t}=1) &=& \varepsilon\exp{\left(-\frac{x^{2}}{4\eta_{\rm O}}\right)}\exp{\left(-\frac{z^{2}}{4\eta_{\rm C}}\right)}, \label{by3d} \\ 
B_{z}(\bar{t}=1) &=& B_{0}+ \varepsilon\exp{\left[-\frac{x^{2}+y^{2}}{4\eta_{C}}\right]}, \label{bz3d}
\label{bequations3dinitial}
\end{eqnarray}
with $\varepsilon = 0.001 B_0$ as before. The analytic temporal evolution of the magnetic field for $\bar{t}>1$ is:
\begin{eqnarray}
B_{x}(\bar{t}) &=& \frac{\varepsilon}{\bar{t}}\exp{\left(-\frac{y^{2}}{4\eta_{\rm O} \bar{t}}\right)}\exp{\left(-\frac{z^{2}}{4\eta_{\rm C}\bar{\bar{t}}}\right)}, \label{bx3dtno0} \\ 
B_{y}(\bar{t})  &=& \frac{\varepsilon}{\bar{t}}\exp{\left(-\frac{x^{2}}{4\eta_{\rm O} \bar{t}}\right)}\exp{\left(-\frac{z^{2}}{4\eta_{\rm C}\bar{t}}\right)}, \label{by3dtno0} \\ 
B_{z}(\bar{t})  &=& B_{0}+ \frac{\varepsilon}{\bar{t}}\exp{\left[-\frac{x^{2}+y^{2}}{4\eta_{C}\bar{t}}\right]}, \label{bz3dtno0}
\label{bequations3d}
\end{eqnarray}

\begin{figure}
\centering
\resizebox{\hsize}{!}{\includegraphics{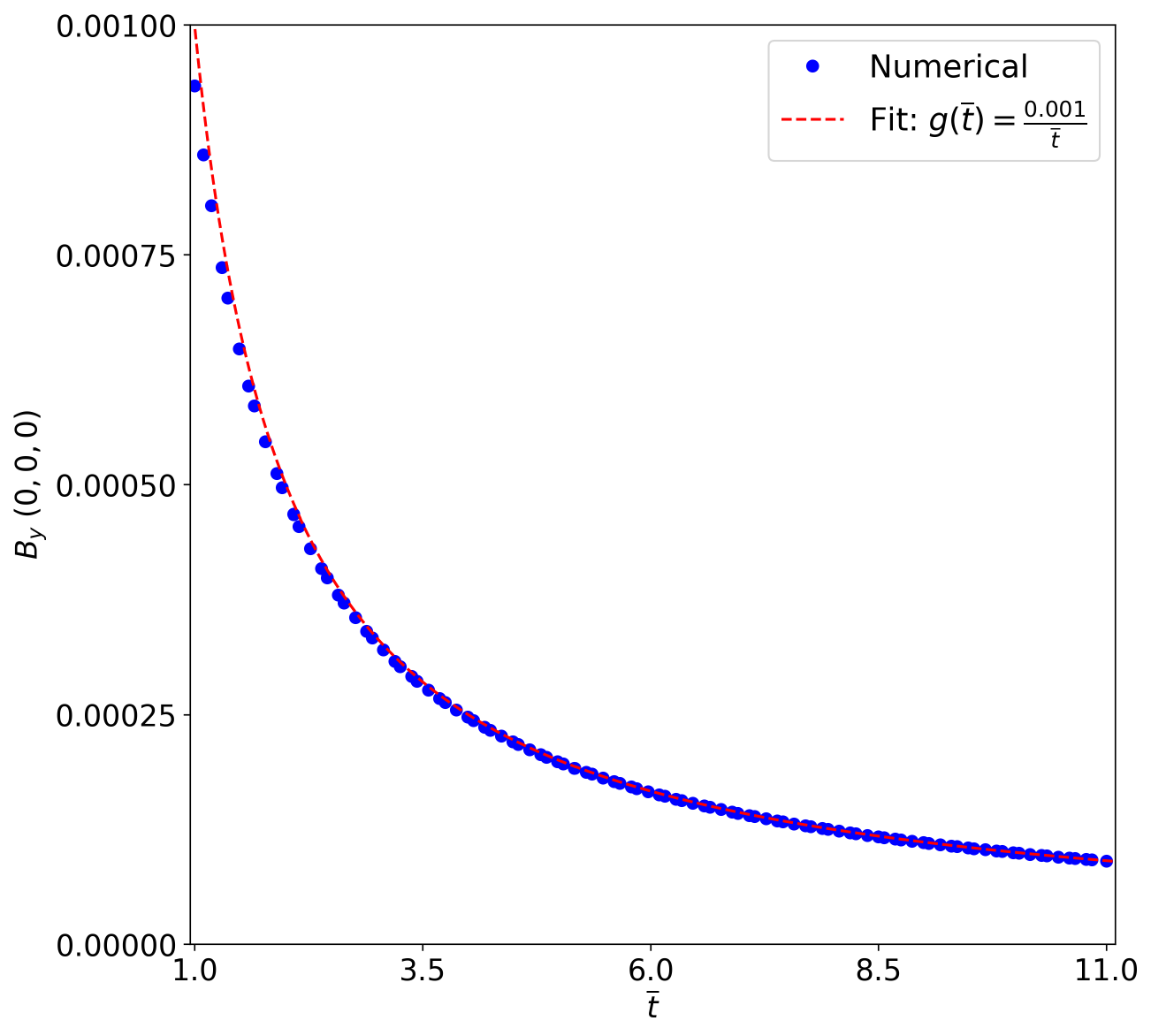}} 
\caption{Temporal evolution of the $y-$component of the magnetic field at the center of the numerical domain. The blue dots corresponds to the results from the 3D simulation while the red dashed line corresponds to the fitting of the data by the function: $ g(\overline{t})=1.0/\overline{t} $. Code units are used.}
\label{Fig_cowling_3D}
\end{figure}

As a verification of the 3D test, we plot in Fig. \ref{Fig_cowling_3D} the evolution of the $y-$component of the magnetic field at the center of the numerical domain, $x=y=z=0$. According to Equation~(\ref{bz2d}), $B_y$ should decrease as $1/\bar{t}$ in that point. As expected, the numerical data follows the theoretical dependence. The other components of the magnetic field are also correctly evolved (not shown here). Therefore, the diffusion problem of the magnetic field is correctly solved in 3D.

\end{appendix}
\end{document}